\documentclass[twocolumn,prb,superscriptaddress,showpacs,showkeys,amsmath,amssymb,pdftex]{revtex4-1}

\usepackage[paperwidth=210mm,paperheight=297mm,centering,hmargin={1.8cm,1.3cm}, vmargin={1.7cm,2.1cm}]{geometry}
\usepackage{graphicx}

\begin{document}

\title{Electronic structure of FeSe monolayer superconductors: shallow bands and correlations}

\author{I.A. Nekrasov}
\email[]{nekrasov@iep.uran.ru}
\affiliation{Institute for Electrophysics, Russian Academy of Sciences, Ural Branch, Amundsen str. 106, Ekaterinburg, 620016, Russia}
\author{N.S. Pavlov}
\email[]{pavlov@iep.uran.ru}
\affiliation{Institute for Electrophysics, Russian Academy of Sciences, Ural Branch, Amundsen str. 106, Ekaterinburg, 620016, Russia}
\author{M.V. Sadovskii}
\email[]{sadovski@iep.uran.ru}
\affiliation{M.N. Mikheev Institute for Metal Physics, Russian Academy of Sciences, Ural Branch, S. Kovalevsky str. 18, Ekaterinburg, 620290, Russia}
\affiliation{Institute for Electrophysics, Russian Academy of Sciences, Ural Branch, Amundsen str. 106, Ekaterinburg, 620016, Russia}


\begin{abstract}

Electronic spectra of typical single FeSe layer 
superconductors -- FeSe monolayer films on SrTiO$_3$ substrate (FeSe/STO) 
and K$_x$Fe$_{2-y}$Se$_{2}$ obtained from ARPES data reveal several puzzles:
what is the origin of shallow and the so called ``replica'' bands near M-point and why the hole-like
Fermi surfaces near $\Gamma$-point are absent. Our extensive LDA+DMFT 
calculations show that correlation effects on Fe-3d states
can almost quantitatively reproduce rather complicated band structure, which is observed in ARPES,
in close vicinity of the Fermi level for FeSe/STO and K$_x$Fe$_{2-y}$Se$_{2}$. Rather unusual shallow electron-like bands
around the M(X)-point in the Brillouin zone are well reproduced.
However, in FeSe/STO correlation effects are apparently insufficient to eliminate the 
hole-like Fermi surfaces around the $\Gamma$-point, which are not 
observed in most ARPES experiments.
Detailed analysis of the theoretical and experimental quasiparticle bands
with respect to their origin and orbital composition is performed.
It is shown that for FeSe/STO system the LDA calculated Fe-3d$_{xy}$ band, renormalized by 
electronic correlations within DMFT gives the quasiparticle band almost
exactly in the energy region of the experimentally 
observed ``replica'' quasiparticle band at the M-point.
For the case of K$_x$Fe$_{2-y}$Se$_{2}$ most bands
observed in ARPES can also be understood as correlation renormalized Fe-3d 
LDA calculated bands, with overall semi-quantitative agreement with our
LDA+DMFT calculations. Thus the shallow bands near the M-point are
common feature for FeSe-based systems, not just FeSe/STO. We also present some simple estimates of ``forward 
scattering'' electron-optical phonon interaction at FeSe/STO interface, 
showing that it is apparently irrelevant for the formation of ``replica'' 
band in this system and significant increase of superconducting $T_c$.

\end{abstract}

\pacs{74.20.-z, 74.20.Rp, 74.25.Jb, 74.70.-b}

\keywords{iron based superconductors; FeSe layered superconductors; DFT/LDA band
structure; LDA+DMFT method; electronic correlations}

\maketitle

\section{Introduction}

The discovery of a class of iron pnictide superconductors has revived
the intensive search and studies of new of  high-temperature superconductors (cf. reviews [\onlinecite{Sad_08, Hoso_09, John, MazKor, Stew, Kord_12}]).
Now there is general agreement that despite many similarities the nature of 
superconductivity in these materials significantly differs from that in 
high -- $T_c$ cuprates, and further studies of these new systems may lead to 
better understanding of the problem of high-temperature superconductivity 
in general.

Actually, the discovery of superconductivity in iron pnictides was very soon 
followed by its discovery in iron {\em chalcogenide} FeSe, which attracted
much interest due to its relative simplicity, though its superconducting 
characteristics (under normal conditions) were rather modest ($T_c\sim$8K). 
Its electronic structure is now well understood and quite similar to that of 
iron pnictides (cf. review in [\onlinecite{FeSe}]).

However, the general situation with iron chalcogenides has changed rather
dramatically with the appearance of {\em intercalated} FeSe based systems 
raising the value of $T_c$ to 30-40K. It was soon recognized that their 
electronic structure is in general rather different form that in iron pnictides
[\onlinecite{JMMM, JTLRev}]. The first system of this kind was A$_x$Fe$_{2-y}$Se$_2$ 
(A=K,Rb,Cs) with $T_c\sim$ 30K [\onlinecite{FeSe_Cryst, AFeSe2}]. It is generally believed 
that superconductivity in this system appears in an ideal 122-type structure,
though most of the samples studied so far were multiphase, consisting of a
mixture of mesoscopic superconducting and insulating (antiferromagnetic)
structures (e.g. such as K$_2$Fe$_4$Se$_5$), complicating the studies of this
system [\onlinecite{Maziopa}].

Further increase of $T_c$ up to 45K has been achieved by
intercalation of FeSe layers with rather large molecules in compounds such as
Li$_x$(C$_2$H$_8$N$_2$)Fe$_{2-y}$Se$_2$ [\onlinecite{intCH}] and
Li$_x$(NH$_2$)$_y$(NH$_3$)$_{1-y}$Fe$_2$Se$_2$ [\onlinecite{intNH}].
The growth of $T_c$ in these systems is sometimes associated with increase of the
distance between the FeSe layers, i.e. with the growth of the two-dimensional
nature of the materials. Recently the active studies has started of
[Li$_{1-x}$Fe$_x$OH]FeSe system with the value of $T_c\sim$43K
[\onlinecite{LiOH1, LiOH2}], where a good enough
single -- phase samples and single crystals were obtained.

A significant breakthrough in the studies of iron chalcogenide superconductors
occurred with the observation of a record high $T_c$ in epitaxial films of single
FeSe monolayer on SrTiO$_3 $(STO) substrate [\onlinecite{FeSe_STO1}].
These films were grown in Ref. [\onlinecite{FeSe_STO1}] and in most of the papers
to follow on the 001 plane of the STO. It should be noted that these films
are very unstable on the air. Thus in many works the resistive transitions were
mainly studied on films covered with amorphous Si or several FeTe layers, which
significantly reduced the observed values of $T_c$. Unique measurements of
the resistance of FeSe films on STO, done in Ref. [\onlinecite{FeSe_STO2}] {\em in situ},
produced the record values of $T_{c} >$100K. However, up to now these results were
not confirmed by independent measurements. Many ARPES measurements of the
temperature behavior of superconducting gap in such films, now confidently
demonstrate the values of $T_c$ in the range of 65--75K, sometimes even higher.

Films consisting of several FeSe layers usually produce the values of $T_c$
much lower than those for the single -- layer films [\onlinecite{FeSe13UCK}].
Monolayer FeSe film on 110 plane of STO  covered with
several FeTe layers was studied in Ref. [\onlinecite{FeSe_STO_FeTe}].
Resistivity measurements (including the measurements of the upper critical
magnetic field $ H_ {c2} $) produced the value of $T_c\sim$30K. FeSe film,
grown on BaTiO$_3$ (BTO) substrate, doped with Nb (with even larger values of
the lattice constant $\sim$ 3.99 \AA), showed (in ARPES measurements) the
value of $T_c\sim $ 70K [\onlinecite{FeSe_BTO}]. In Ref. [\onlinecite{FeSe_Anatas}]
quite high values of the superconducting gap were reported (from
tunneling spectroscopy) for FeSe monolayers grown on 001 plane of TiO$_2 $
(anatase), which in its turn was grown on the 001 plane of SrTiO$_3$.
The lattice constant of anatase is actually very close to the lattice constant
of bulk FeSe, so these FeSe film were essentially unstretched.

Single -- layer FeSe films were also grown on the graphene substrate, but the
value of $T_c$ obtained was of the order of 8-10K as in bulk FeSe [\onlinecite{FeSeGraph}].
This emphasizes the possible unique role of substrates such as
Sr(Ba)TiO$_3$ in the significant increase of $T_c$.

More information on FeSe/STO films and other monolayer FeSe systems can be found
in recent reviews [\onlinecite{Maziopa,FeSe1UC_rev,Sad_16}].

\section{Crystal structures of iron based superconductors}

\begin{figure*}
\center{\includegraphics[clip=true,width=0.7\textwidth]{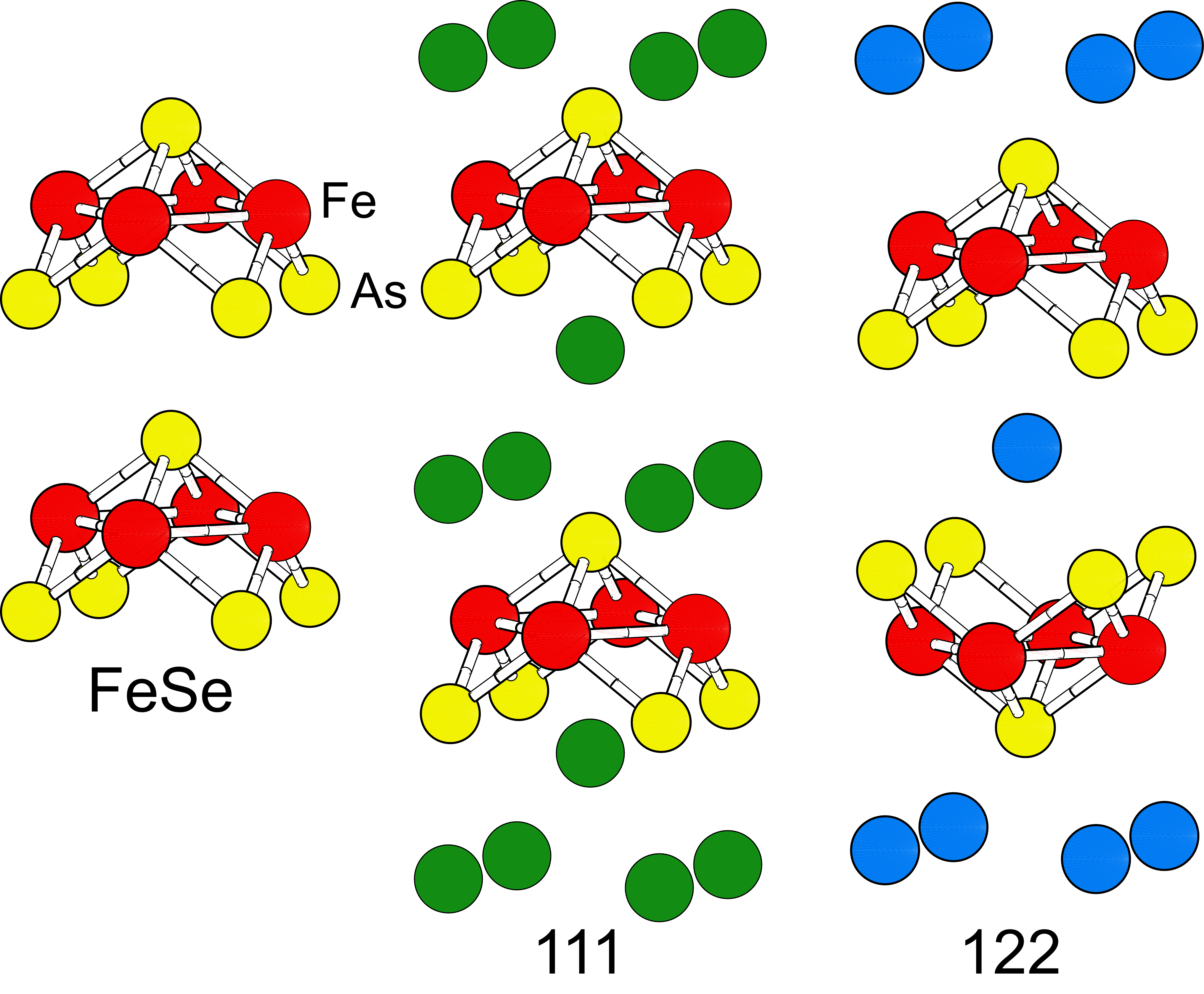}}
\caption{Typical crystal structures of iron based superconductors.}
\label{StrucFePn}
\end{figure*}

In Figure \ref{StrucFePn} we schematically show the simple crystal
structure of typical iron based superconductors
[\onlinecite{ FeSe, Sad_08, Hoso_09, John, MazKor, Stew, Kord_12}]. The common element
here is the FeAs or FeSe planes (layers), with Fe ions forming a simple square
lattice. The pnictogen (Pn - As) or chalcogen (Ch - Se) ions here
are at the centers of the Fe squares above and below Fe plane.
The 3d states of Fe in FePn plane (Ch) are decisive in the
formation of the electronic structure of these systems, determining
superconductivity. In a sense, these layers are quite similar to the CuO$_2$
planes of cuprates (copper oxides) and these systems can also be considered
approximately as quasi-two dimensional conductors.

Note that all of the FeAs crystal structures shown in Fig. \ref{StrucFePn}
are ion--covalent crystals. Chemical formula, say for a typical 122 system, can
be written for example as Ba$^{+2}$(Fe$^{+2}$)$_2$(As$^{-3}$)$_2$. Here the
charged FeAs layers are held by Coulomb forces from the surrounding ions.
In the bulk FeSe electrically neutral FeSe layers are connected to each other
by much weaker van der Waals interactions. This makes FeSe system most suitable
for intercalation by various atoms and molecules that can be fairly easy
introduced between the layers of FeSe. Chemistry of intercalation processes for
iron chalcogenide superconductors is discussed in detail in a recent review of
Ref. [\onlinecite{VivRod}].

\begin{figure}
\center{\includegraphics[clip=true,width=0.25\textwidth]{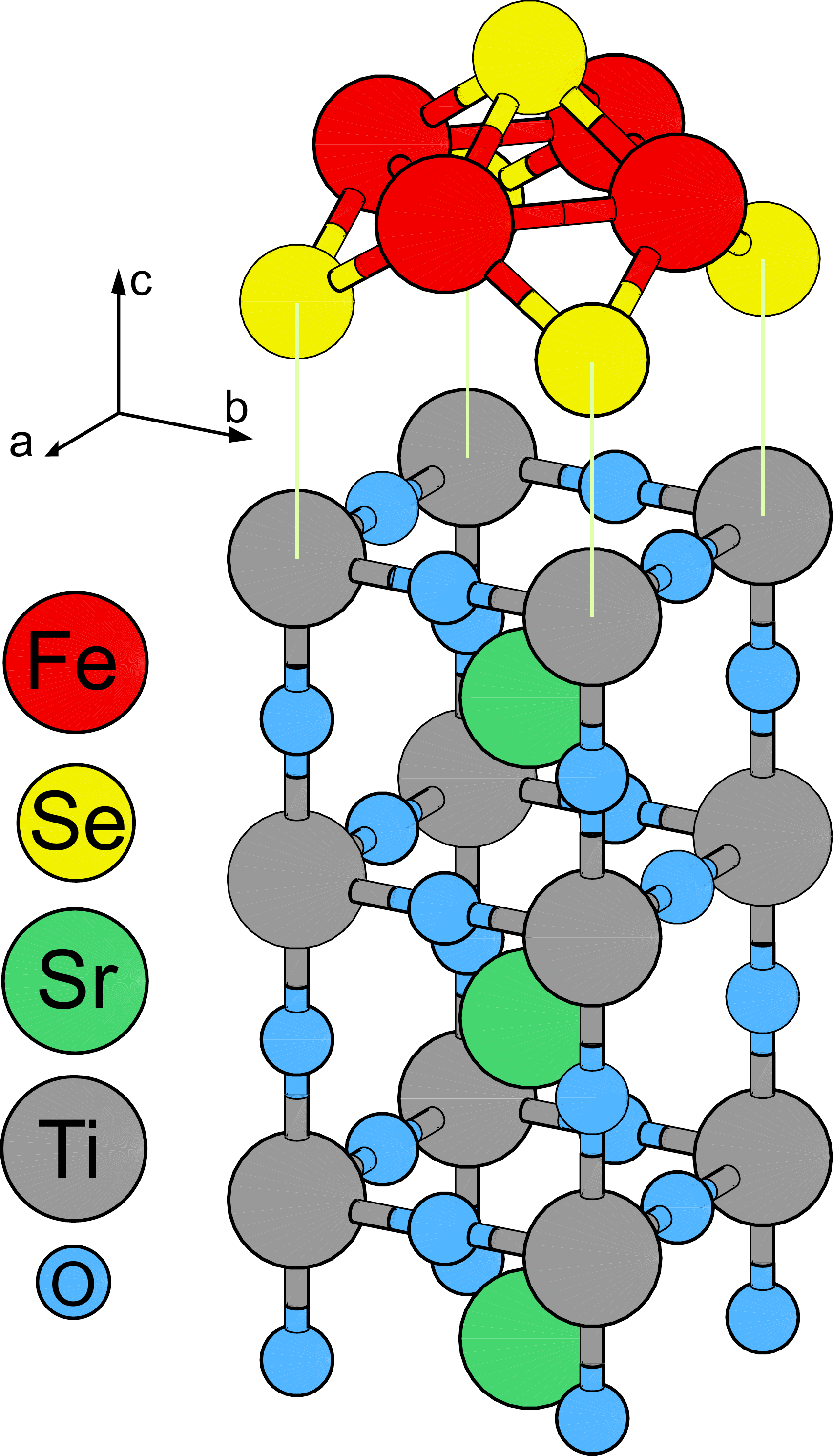}}
\caption{Crystal structure of FeSe monolayer on (001)  surface of SrTiO$_3$with TiO$_2$ topmost layer.}
\label{FeSeSTO}
\end{figure}

\subsection{FeSe, FeSe/STO}

Bulk FeSe system has probably the simplest crystal structure among iron
high-T$_c$ superconductors. It has tetragonal structure with the space
group $P$4/$nmm$ and lattice parameters $a=3.765$~\AA, $c=5.518$~\AA.
The experimentally observed crystallographic positions are: Fe(2a)
(0.0, 0.0, 0.0), Se(2c) (0.0, 0.5, z$_{Se}$), z$_{Se}$=0.2343~[\onlinecite{param}].
In our LDA calculations of isolated FeSe layer the slab technique was used with
these crystallographic parameters.

The FeSe/STO crystal structure was taken from LDA calculation with crystal
structure relaxation~[\onlinecite{FeSe_STO_param}]. In slab approach FeSe monolayer was
placed on three TiO$_2$-SrO layers to model the bulk SrTiO$_3$ substrate.
The FeSe/STO slab crystal structure parameters used were $a=3.901$~\AA, Ti-Se
distance $3.13$~\AA, Fe-O distance $4.43$~\AA,  distance between  top (bottom)
Se ion and the Fe ions plane is 1.41~\AA\ (1.3~\AA). Atomic positions used were:
Sr -- (0.5,0.5,-1.95~\AA), O -- (0.5,0,0), (0,0,-1.95~\AA), Ti -- (0,0,0).

The structure of the FeSe monolayer film on STO is shown in Fig. \ref{FeSeSTO}.
Here the FeSe layer is directly adjacent to the surface TiO$_2$ layer of STO.
The lattice constant within FeSe layer in a bulk samples is equal to 3.77\AA,
while STO has substantially greater lattice constant equal to 3.905 \AA, so
that the single -- layer FeSe film should be noticeably stretched, as
compared with the bulk FeSe. However this tension quickly disappears as the
number of subsequent layers grows.

\subsection{KFe$_2$Se$_2$}

\begin{figure}
\center{\includegraphics[clip=true,width=0.256\textwidth]{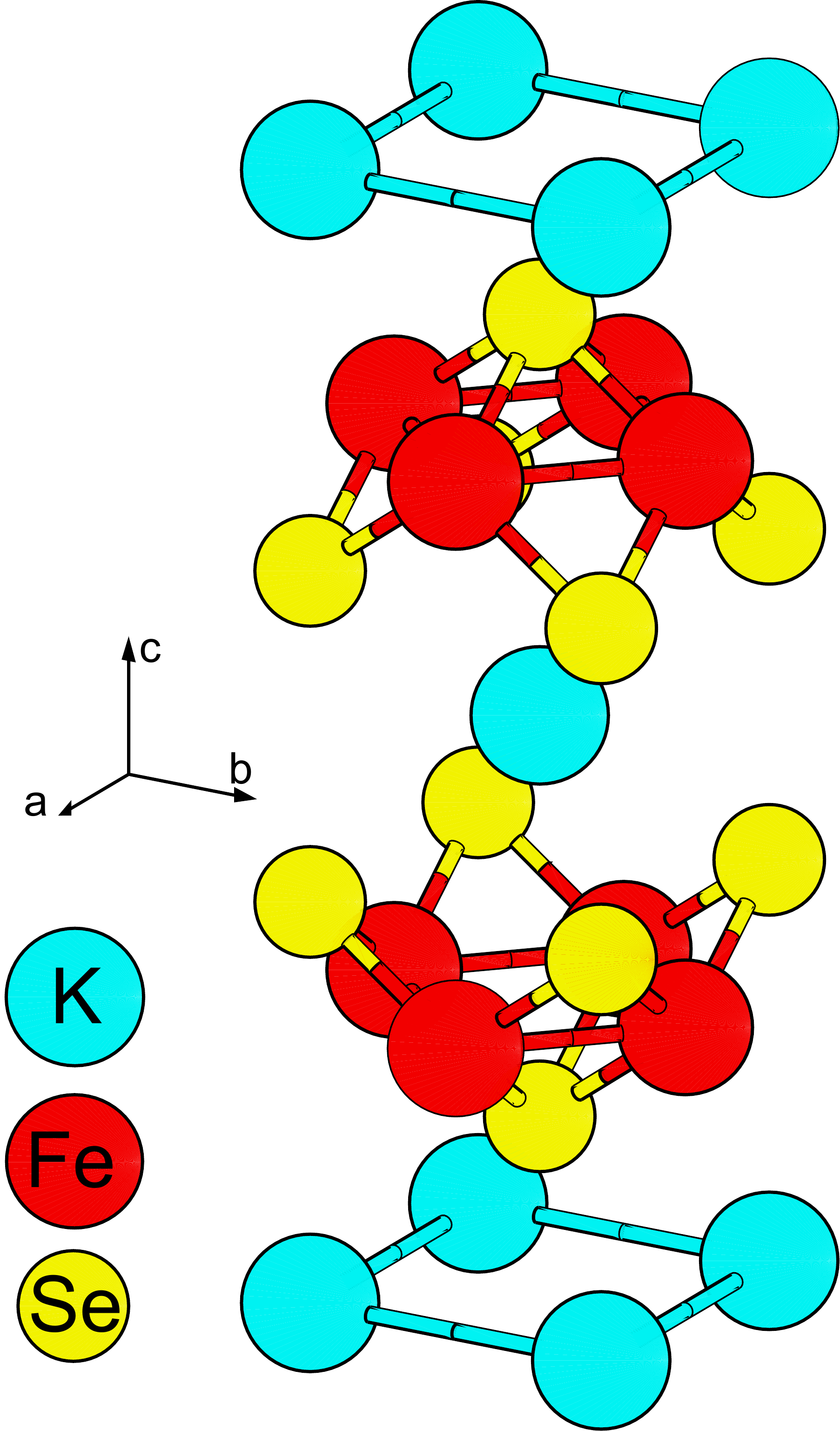}}
\caption{Ideal ($x$=1) crystal structure
of 122-type of K$_x$Fe$_2$Se$_2$ compound.}
\label{interFeCh}
\end{figure}

The ideal KFe$_2$Se$_2$ compound has tetragonal structure with the space group
$I$4/$mmm$ and lattice parameters $a=3.9136$~\AA~and $c=14.0367$~\AA.
The crystallographic positions are: K(2a)
(0.0, 0.0, 0.0), Fe(4d) (0.0, 0.5, 0.25), Se(4e) (0.0, 0.5, z$_{Se}$) with
z$_{Se}$=0.3539~[\onlinecite{FeSe_Cryst}]. The crystal structure of K$_x$Fe$_2$Se$_2$
is shown in Fig. \ref{interFeCh}.

Below we compare the ARPES detected quasiparticle bands for
FeSe/STO and K$_x$Fe$_{2-y}$Se$_{2}$ and make comparison of these bands with
the results of LDA+DMFT calculations for these systems, as well as for isolated
FeSe layer, together with the  analysis of initial LDA calculated
bands~[\onlinecite{Nekrasov_FeSe}].

\section{Computation details}

Our LDA$'$ calculations~[\onlinecite{CLDA,CLDA_long}] of KFe$_2$Se$_2$ compound were
performed using the Linearized Muffin-Tin Orbitals method (LMTO)~[\onlinecite{LMTO1,LMTO2,LMTO3}].
The electronic structures of FeSe monolayer and FeSe monolayer on SrTiO$_3$
substrate were calculated within FP-LAPW method~[\onlinecite{wien2k}].

For DMFT part of LDA+DMFT calculations we always employed the CT-QMC impurity
solver~[\onlinecite{ctqmc1,ctqmc2,ctqmc3,triqs}]. To define DMFT lattice problem for KFe$_2$Se$_2$
compound we used the full LDA Hamiltonian,  same as in
Refs.~[\onlinecite{KFeSeLDADMFT1,KFeSeLDADMFT2}].
For isolated FeSe layer and FeSe/STO projection on Wannier functions was done
for Fe-3d and Se-4p states (isolated FeSe layer) and for Fe-3d, Se-4p states and
O-2p$_y$ states from TiO$_2$ layer adjacent to SrTiO$_3$ (FeSe/STO).
To this end the standard  wien2wannier interface~[\onlinecite{wien2wannier}] and
wannier90 projecting technique~[\onlinecite{wannier90}] were applied.

The DMFT(CT-QMC) computations were done at reciprocal temperature
$\beta=$40 eV$^{-1}$ ($\sim$290 K) with about 10$^8$ Monte-Carlo sweeps.
Interaction parameters of Hubbard model were taken $U$=5.0 eV, $J$=0.9 eV for
isolated FeSe and FeSe/STO and $U$=3.75 eV, $J$=0.56 eV for
KFe$_2$Se$_2$~[\onlinecite{KFe2Se2_ARPES}]. We employed the self-consistent
fully-localized limit definition of the double-counting
correction~[\onlinecite{CLDA_long}].
Thus computed values of Fe-3d occupancies and corresponding double-counting
energies are $E_{dc}=18.886$ eV, $n_d=5.79$ (K$_{0.76}$Fe$_{1.72}$Se$_2$),
$E_{dc}=31.63$ eV, $n_d=7.35$ (isolated FeSe layer), $E_{dc}=30.77$ eV, $n_d=7.16$
LDA+DMFT calculations of FeSe/STO were performed for doping level of 0.2
electrons per Fe ion.
Chemical composition K$_{0.76}$Fe$_{1.72}$Se$_2$ corresponds to the total number
of electrons 26.52 per unit cell which corresponds to the doping level of
1.24 holes per Fe ion. This doping level was taken for LDA$'$+DMFT calculations.
Moreover a number of LDA+DMFT calculations of FeSe/STO for various model parameters
can be seen in the Appendix.

The LDA+DMFT spectral function maps were obtained after analytic continuation
of the local self-energy $\Sigma(\omega)$ from Matsubara to real frequencies.
To this end we have used the Pade approximant algorithm~[\onlinecite{pade}] and
checked the results with the maximum entropy method~[\onlinecite{ME}] for Green's
function G($\tau$).

\section{Electronic structure of iron -- selenium systems}

Electronic spectrum of iron pnictides now is well understood, both from
theoretical calculations based on the modern band structure theory and ARPES
experiments [\onlinecite{Sad_08, Hoso_09, John, MazKor, Stew, Kord_12}]. It is clear
that almost all physics related to superconductivity is determined by
electronic states of FeAs plane (layer), shown in Fig. \ref{StrucFePn}.
The spectrum of carriers in the vicinity of the Fermi level $\pm$ 0.5 eV, where
superconductivity is formed, practically have only Fe-3$d$ character.
The Fermi level is crossed by up to five bands (two or three hole and two
electronic ones), forming a typical spectrum of a semi -- metal.

In this rather narrow energy interval near the Fermi level this dispersions can
be considered as parabolic [\onlinecite{MazKor,KuchSad10}]. Most LDA+DMFT calculations
[\onlinecite{DMFT1, DMFT2}] show that the role of electronic correlations
in iron pnictides, unlike in the cuprates, is relatively insignificant. It is
reduced to more or less significant effective mass renormalization of the
electron and hole dispersions, as well as to general narrowing (``compression'')
of the bandwidths.

The presence of the electron and hole Fermi surfaces of similar size, satisfying
(approximately) the ``nesting'' condition plays an important role in the
theories of superconducting pairing in iron arsenides based on (antiferromagnetic)
spin fluctuation mechanism of pairing [\onlinecite{MazKor}]. We shall see below that the
electronic spectrum and Fermi surfaces in the Fe chalcogenides are very
different from those in Fe pnictides. This raises the new problems for the
understanding of microscopic mechanism of superconductivity in FeSe systems.

\subsection{A$_x$Fe$_2$Se$_2$  system}

\subsubsection{DFT/LDA results}

\begin{figure*}
\center{\includegraphics[clip=true,width=0.45\textwidth]{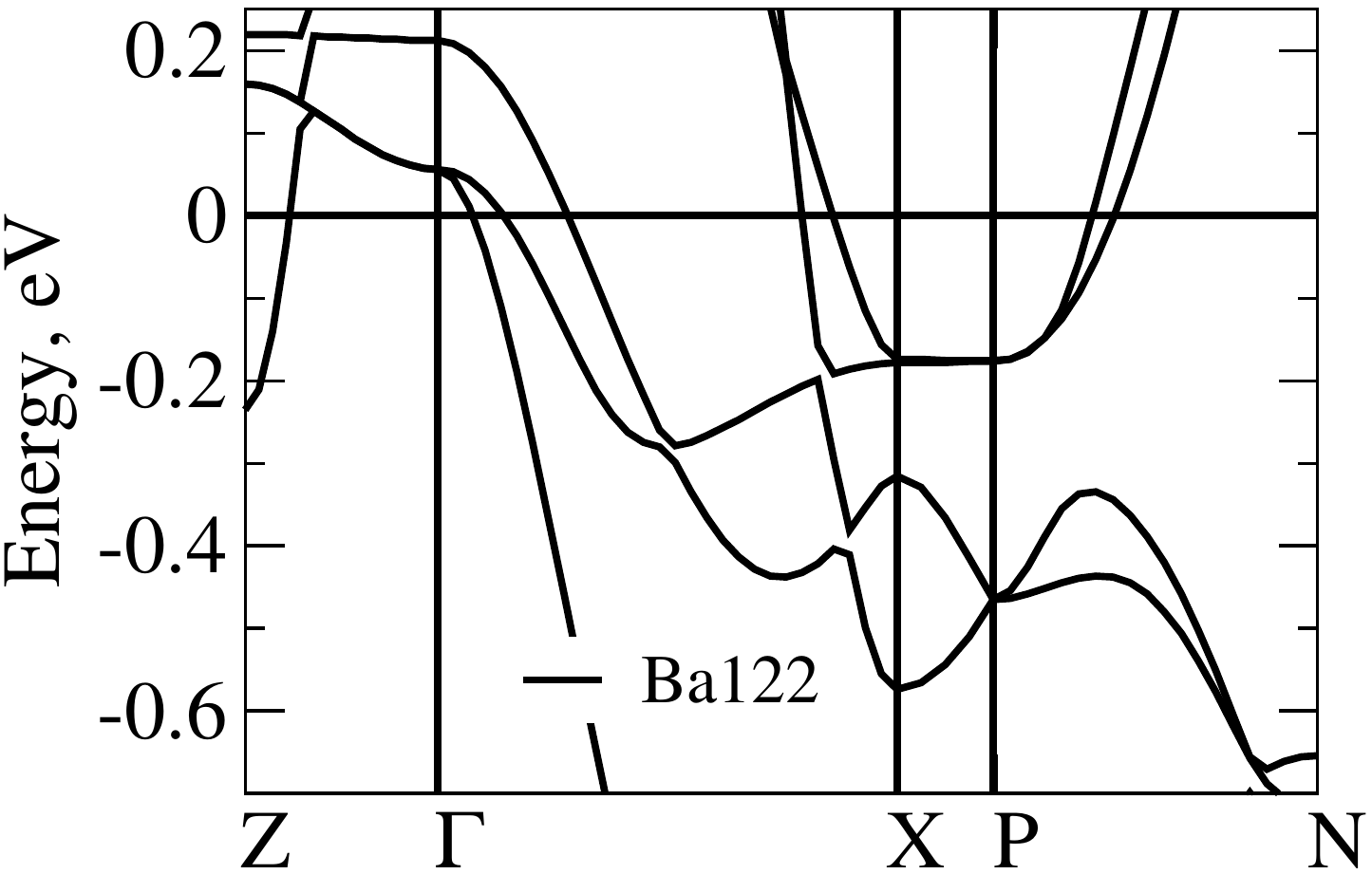}
        \includegraphics[clip=true,width=0.45\textwidth]{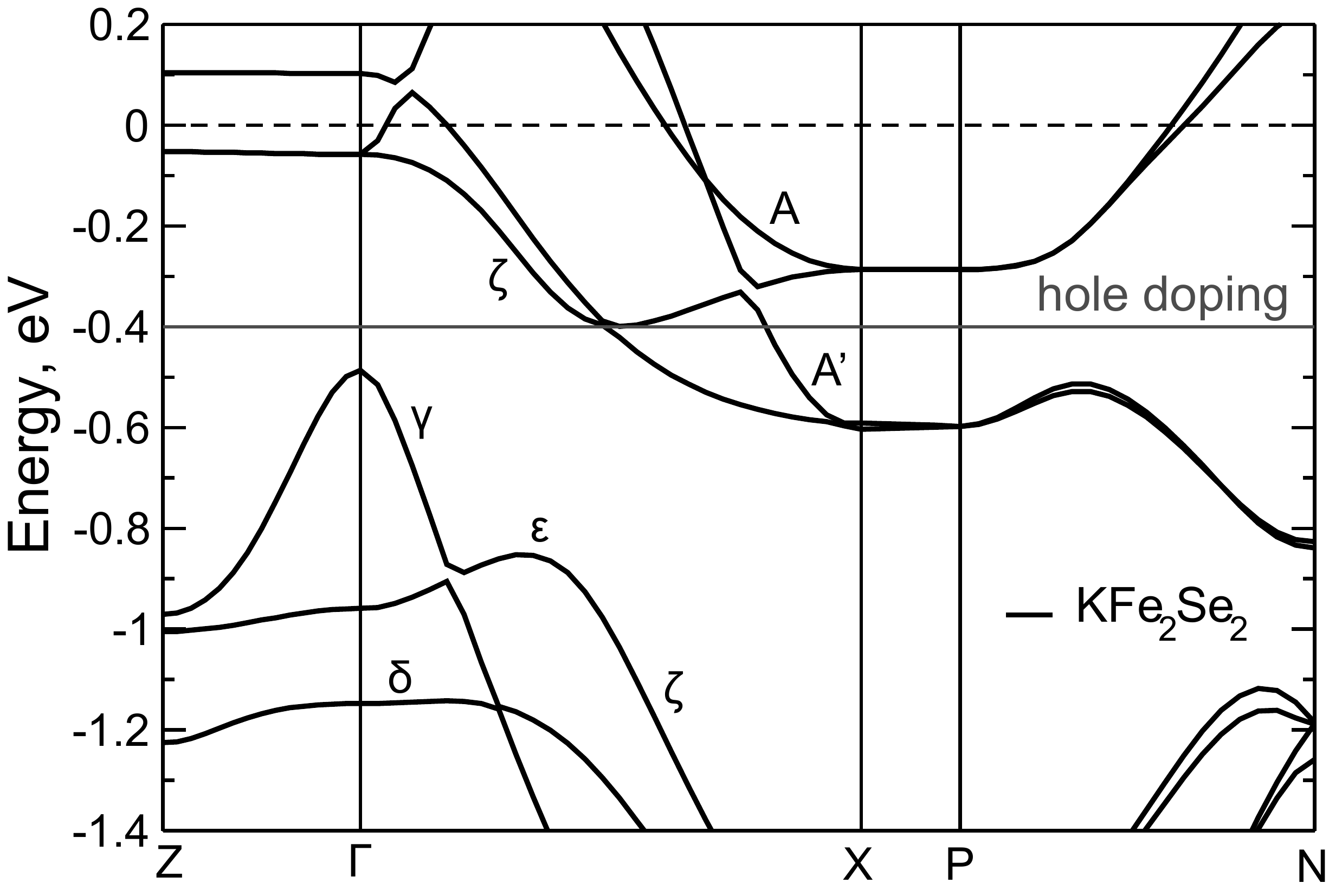}}
\caption{Left -- LDA bands of Ba122 near the Fermi level ($E=$0) [\onlinecite{Ba122}],
right -- LDA$'$ bands of KFe$_2$Se$_2$. Additional horizontal lines correspond to
Fermi level of  60\% hole doping [\onlinecite{KFe2Se2}]. The letters designate bands in
the same way as in Fig.~\ref{fig1sto}.}
\label{122comp}
\end{figure*}

First LDA calculations of electronic structure of the A$_x$Fe$_{2-y}$Se$_2$
(A=K,Cs) system were performed soon after its experimental discovery
[\onlinecite{KFe2Se2, KFe2Se2SI}]. Surprisingly enough, this spectrum was discovered to
be qualitatively different from that of the bulk FeSe and spectra of
practically all known systems based on FeAs. In Fig. \ref{122comp} on the left we show
energy bands of BaFe$_2$As$_2 $ (Ba122) [\onlinecite{Ba122}] (which is the typical
prototype of all FeAs systems) and those of KFe$_{2}$Se$_2$ [\onlinecite{KFe2Se2}] on the right.
One can observe a significant difference in the spectra near the Fermi level.

\begin{figure*}
\center{\includegraphics[clip=true,width=0.9\textwidth]{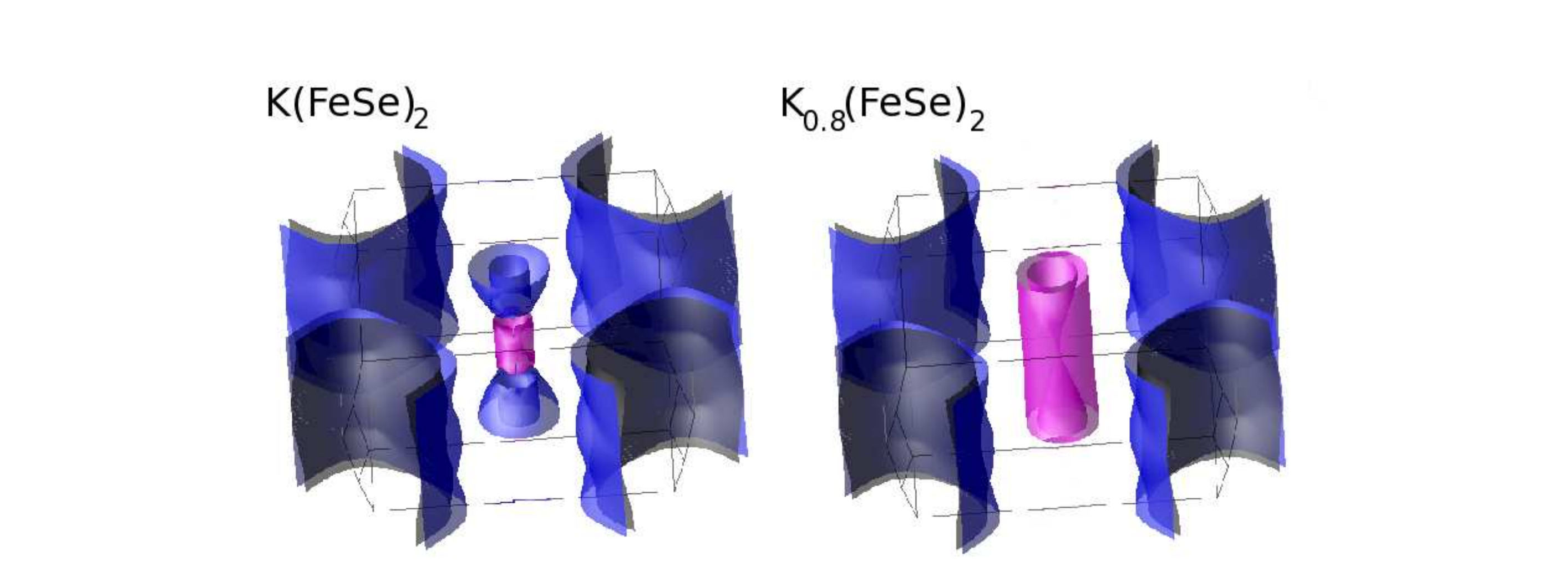}}
\caption{LDA Fermi surfaces for the stoichiometric KFe$_2$Se$_2$
(left) and at 20\% hole doping (right)  [\onlinecite{KFe2Se2}]. }
\label{FSAFeSe}
\end{figure*}

In Fig. \ref{FSAFeSe} we show the calculated  Fermi surfaces for
K$_x$Fe$_{2-y}$Se$_2$ system at various doping levels [\onlinecite{KFe2Se2}].
These differ significantly from the Fermi surfaces of FeAs systems
--- in the center of the Brillouin zone, there are only small
Fermi sheets of electronic nature, while the electronic cylinders in
the Brillouin zone corners are substantially larger.
The shapes of the Fermi surfaces, typical for bulk FeSe and FeAs systems,
can be obtained only at a much larger (apparently experimentally inaccessible)
levels of the hole doping [\onlinecite{KFe2Se2}].

\begin{figure*}
\center{\includegraphics[clip=true,width=0.75\textwidth]{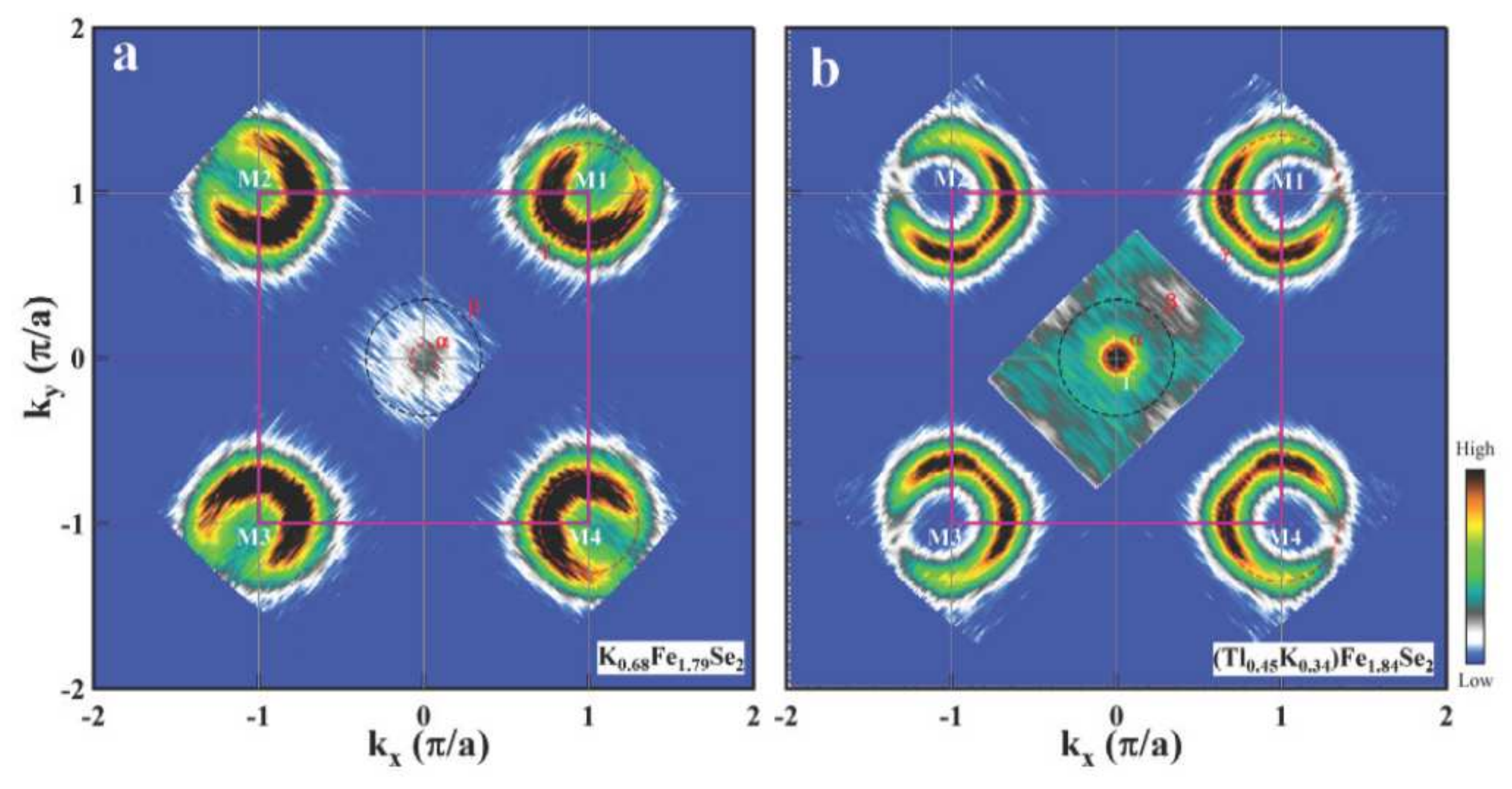}}
\caption{ARPES Fermi surfaces
of K$_{0.68}$Fe$_{1.79}$Se$_2$
($T_c$=32K) and  Tl$_{0.45}$K$_{0.34}$Fe$_{1.84}$Se$_2$ ($T_c$=28K)
[\onlinecite{KFe2Se2ARPES}].}
\label{FSFeSeARP}
\end{figure*}

This shape of the Fermi surfaces in K$_x$Fe$_{2-y}$Se$_2$ systems
was almost immediately confirmed in ARPES experiments.
For example, in Fig. \ref{FSFeSeARP} we show the ARPES data of Ref.
[\onlinecite{KFe2Se2ARPES}], which are obviously in qualitative agreement with LDA
results of Refs. [\onlinecite{KFe2Se2,KFe2Se2SI}].
Note, that in this system it is clearly impossible to speak of any, even
approximate, ``nesting'' properties of electron and hole Fermi surfaces.

\subsubsection{LDA+DMFT results}

LDA+DMFT and LDA$'$+DMFT calculations for K$_{1-x}$Fe$_{2-y}$Se$_2$ system for
various doping levels were performed in  Refs.
[\onlinecite{KFeSeLDADMFT1,KFeSeLDADMFT2,FeSe_STO_2017}].
The results of these calculations can be directly compared with
the ARPES data obtained in Refs. [\onlinecite{KFe2Se2_ARPES,KFe2Se2_ARPES_2,NaFeSe}].

\begin{figure}[!ht]
\center{\includegraphics[width=.5\linewidth]{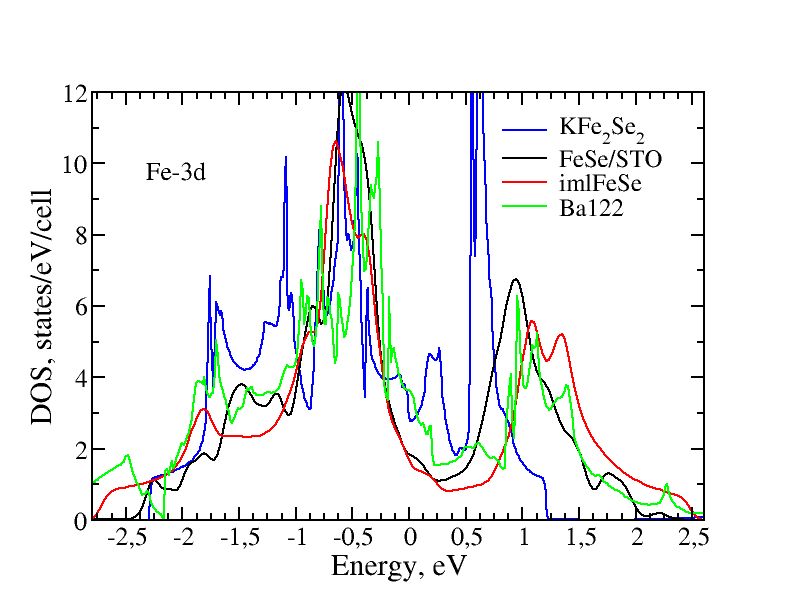}}
\caption{LDA calculated densities of states (DOS) for various iron-based superconductors:
imlFeSe (red, $W$=5.2 eV), Ba122 (green, $W$=4.8 eV), FeSe/STO (black, $W$=4.3 eV), KFe$_{2}$Se$_2$ (blue, $W$=3.5 eV)
Fermi level is at zero energy.}
\label{DOScomp}
\end{figure}

It turns out that in K$_{1-x}$Fe$_{2-y}$Se$_2$ correlation effects are
quite important, leading to a noticeable change of LDA calculated dispersions.
In contrast to iron arsenides, where the quasiparticle bands near the Fermi level
are well defined, in the K$_{1-x}$Fe$_{2-y}$Se$_2$ compounds in the vicinity of
the Fermi level we observe much stronger suppression of the intensity of
quasiparticle bands. This reflects the stronger role of correlations in this
system, as compared to iron arsenides. The value of the quasiparticle
renormalization (correlation narrowing) of the bands at the Fermi level is 4-5,
whereas in iron arsenides this factor is only 2-3 for the same values of the
interaction parameters. That can be understood in terms of $W$ -- width of bare LDA Fe-3d
states. As it is shown on Fig. \ref{DOScomp} the largest bandwidth  $W$=5.2 eV has isolated FeSe
monolayer (red curve), then comes Ba122 (green curve) with $W$=4.8 eV, FeSe/STO (black curve) with $W$=4.3 eV. and
finally the most narrow bare band has KFe$_{2}$Se$_2$ system (blue curve) -- $W$=3.5 eV. In its turn
such lowering of the $W$ can be explained by the growth of lattice constant from imlFeSe to KFe$_{2}$Se$_2$.

The results of these calculations, in general, are in good qualitative agreement
with the ARPES data [\onlinecite{KFe2Se2_ARPES, KFe2Se2_ARPES_2}], which
demonstrate strong damping of quasiparticles in the immediate vicinity of
the Fermi level and a strong renormalization of the effective masses as compared
to systems based on FeAs.

\begin{figure*}[!ht]
\center{\includegraphics[width=.85\linewidth]{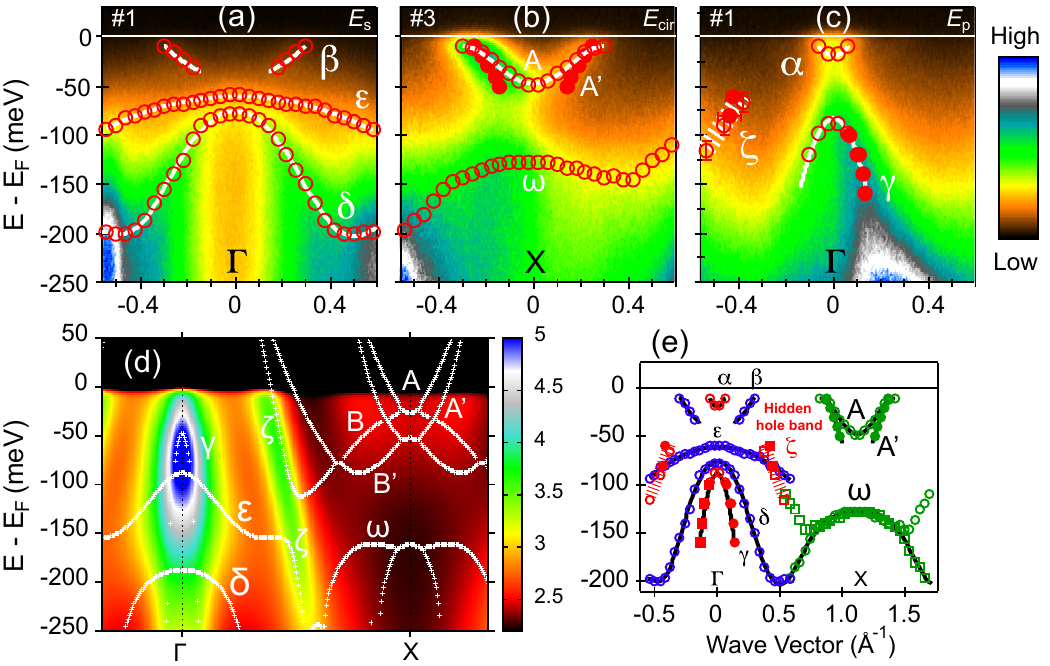}}
\caption{Panels -- (a),(b),(c) ARPES data around $\Gamma$ and X-points
for K$_{0.62}$Fe$_{1.7}$Se$_2$~[\onlinecite{NaFeSe}];
(d) -- LDA$'$+DMFT spectral function map with maxima shown by white crosses
for K$_{0.76}$Fe$_{1.72}$Se$_2$~[\onlinecite{FeSe_STO_2017}];
(e) -- quasiparticle bands extracted from ARPES~[\onlinecite{NaFeSe}].
Bands of similar orbital character are marked with
Greek letters on all panels. Fermi level is at zero energy.}
\label{fig_kfese}
\end{figure*}

In Fig.~\ref{fig_kfese} we present the comparison of LDA+DMFT spectral function
maps (panel (d)) [\onlinecite{FeSe_STO_2017}] and ARPES data of Ref.~[\onlinecite{NaFeSe}]
(panels (a,b,c,e)) for K$_x$Fe$_{2-y}$Se$_{2}$.
Panels (a,b,c) of Fig.~\ref{fig_kfese} correspond to different incident beam
polarizations in ARPES experiment:
$E_s$ -- polarization in the plane parallel to the sample surface;
$E_p$ -- polarization in the plane normal to the sample surface;
$E_{cir}$ -- circular polarization. The use of different polarizations
allows one to distinguish contributions of bands with different symmetry
(see discussion in Refs.~[\onlinecite{NaFeSe,DMFT2}]). This fact is clear from
the data on panels (a,b,c) of Fig.~\ref{fig_kfese} where different bands are
marked with Greek letters.
In Fig.~\ref{fig_kfese}(e) we show the joint picture of all quasiparticle bands
detected in ARPES~[\onlinecite{NaFeSe}] experiment.

Now we can explain the origin of the experimental bands and their
orbital composition on the basis of
LDA$'$~[\onlinecite{CLDA,CLDA_long,KFeSeLDADMFT1,KFeSeLDADMFT2}] calculations
for KFe$_2$Se$_2$ (Fig.~\ref{122comp}, right panel) and LDA$'$+DMFT results
of Ref. [\onlinecite{FeSe_STO_2017}] (Fig.~\ref{fig_kfese}, panel (e)).
In our LDA$'$+DMFT calculations the $A$ quasiparticle band near X-point
corresponds to Fe-3d$_{xz}$ and Fe-3d$_{yz}$ states and the $A'$ quasiparticle
band near X-point is mainly formed by Fe-3d$_{xy}$ states.
These bands are denoted in the same way as on right panel of Fig.~\ref{122comp}.
At about -0.15 eV at the X-point there is $\omega$ quasiparticle band which is
formed in our calculations due to self-energy effects only.

Thus the $A$ band is located at $\sim$ 50 meV where shallow band, typical for
FeSe monolayer materials, is observed experimentally.
Another shallow band $A'$ near M-point has energy about 75 meV. The $A'$ band might be strongly suppressed in the experiments
due to its Fe-3d$_{xy}$ symmetry as it is stressed by the authors of Ref.~[\onlinecite{NaFeSe}].
So both $A$ and $A'$ bands are just correlation renormalized LDA$'$ bands (compare with right panel of
Fig.~\ref{122comp}). Similar conclusion can be given concerning the $B$ band and $B'$ bands.
Also one should note here that quasiparticle masses of the $A$ and $A'$ bands are only slightly
 different. It is also important to say that $A'$ is well defined near the Fermi level and is almost
 undetected near X-point (Fig.~\ref{fig_kfese}, panel (e)).Thus for the case of KFe$_2$Se$_2$ system
 we have evidently shown that purely electronic shallow $A$ and $A'$ bands agree rather well with
 ARPES data~[\onlinecite{NaFeSe}]. Moreover as we will point out later
 FeSe/STO bands near M-point are practically the same. So one can clearly see that
 such  $A$ and $A'$ band dispersions are a common feature of FeSe-based materials and
 probably can be resolved completely in future ARPES experiments.

Let us turn to bands around $\Gamma$-point. The $\varepsilon$ and $\delta$
bands are formed by Fe-3d$_{3z^2-r^2}$ states.
The $\varepsilon$ band is rather strongly modified in comparison
with the initial LDA$'$ $\varepsilon$ band (see Fig.~\ref{122comp}, right panel),
while the $\delta$ band more or less preserves its initial form.
Energy location of $\varepsilon$ quasiparticle band agrees well in LDA$'$+DMFT
and ARPES. However, the $\delta$ band is much lower in energy
in LDA$'$+DMFT calculations. At the $\Gamma$-point the $\gamma$ band
(which is the hybrid band of Fe-3d$_{xz}$, Fe-3d$_{yz}$  and Fe-3d$_{xy}$
states) in LDA$'$+DMFT is above the $\varepsilon$ and $\delta$ bands in
contrast with ARPES data (Fig.~\ref{fig_kfese}(e)). This picture is somehow
inherited from the initial LDA$'$ band structure (Fig.~\ref{122comp}, right).
The $\zeta$ band (Fig.~\ref{fig_kfese}(e)) consists in fact of two bands.
The upper part (above 130 meV) of this band is formed by Fe-3d$_{xz}$ and
Fe-3d$_{yz}$ states, while its lower part is formed by Fe-3d$_{3z^2-r^2}$
states. In ARPES experiments this band is only partially observed around 80 meV
(Fig.~\ref{fig_kfese}(e)), while its lower part is not distinguished
experimentally from $\omega$ band~[\onlinecite{NaFeSe}].

\begin{figure*}[!ht]
\center{\includegraphics[width=.75\linewidth]{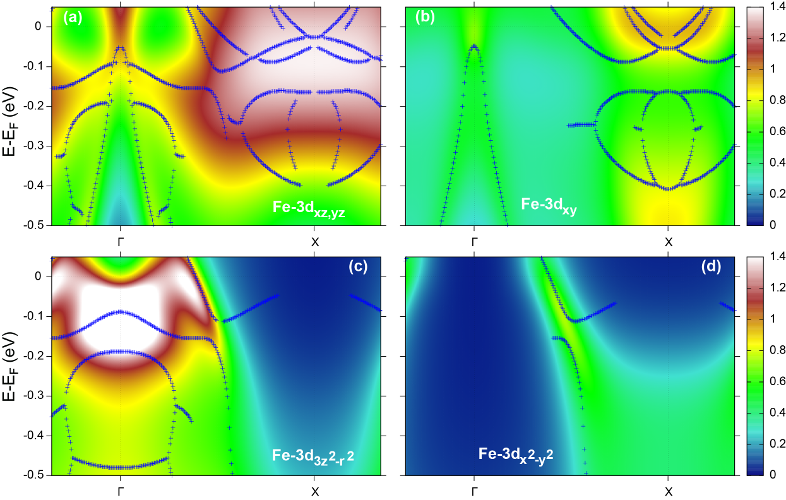}}
\caption{LDA$'$+DMFT spectral function map [\onlinecite{FeSe_STO_2017}]
for different Fe-3d orbitals of
K$_{0.76}$Fe$_{1.72}$Se$_2$: (a) -- Fe-3d$_{xz}$ and Fe-3d$_{yz}$,
(b) -- Fe-3d$_{xy}$, (c) -- Fe-3d$_{3z^2-r^2}$, (d) -- Fe-3d$_{x^2-y^2}$.
Maxima of the spectral density are shown with crosses.
Fermi level is zero energy.}
\label{or_bands_kfese}
\end{figure*}

The orbitally resolved spectral function of K$_{0.76}$Fe$_{1.72}$Se$_2$ is shown
in Fig.~\ref{or_bands_kfese}. Here the bands are rather strongly renormalized by
correlations not only by the constant scaling factor, but also because of band
shapes modifications in comparison to LDA bands. Since electronic correlations
are quite strong for K$_{0.76}$Fe$_{1.72}$Se$_2$ (because of most narrow Fe-3d bare band among considered systems,
see Fig. \ref{DOScomp}) and bands are rather broadened
by lifetime effects we explicitly show here the spectral function maxima
positions by crosses. We can conclude that quasiparticle bands structures around
the Fermi level for both compositions under discussion are rather similar.

The overall agreement between ARPES and LDA$'$+DMFT results for
K$_{0.76}$Fe$_{1.72}$Se$_2$ system is rather satisfactory and allows one to
identify the orbital composition of different bands detected in the experiment.
However $\alpha$ and $\beta$ bands found in ARPES are not observed in our
LDA$'$+DMFT calculated spectral function maps. More so there are no obvious
candidates for these bands within the LDA$'$ band structure (Fig.~\ref{122comp},
right). Thus the origin of experimentally observed $\alpha$ and $\beta$
quasiparticle bands (if there are some) remains at present unclear.

\subsection{FeSe monolayer films}

\subsubsection{DFT/LDA results}

\begin{figure}[!ht]
\center{\includegraphics[width=.45\textwidth]{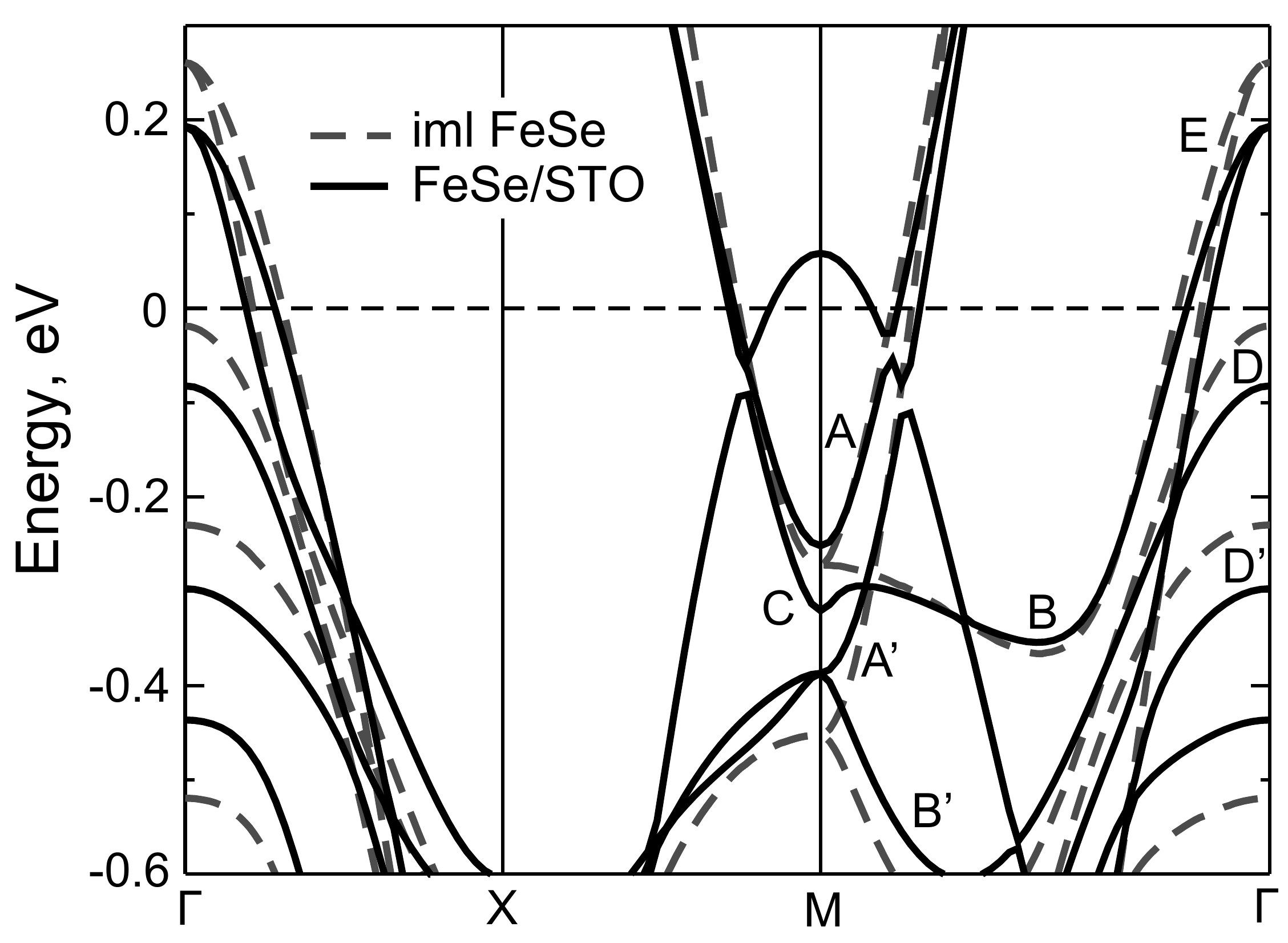}}
\caption{LDA band dispersions of paramagnetic isolated FeSe monolayer (dashed line)
and paramagnetic FeSe/STO (solid line). The letters designate bands in
the same way as in Fig.~\ref{fig1sto}. The Fermi level $E_F$ is at zero energy.}
\label{fese_sto_bands}
\end{figure}

\begin{figure}[!ht]
\center{\includegraphics[clip=true,width=0.4\textwidth]{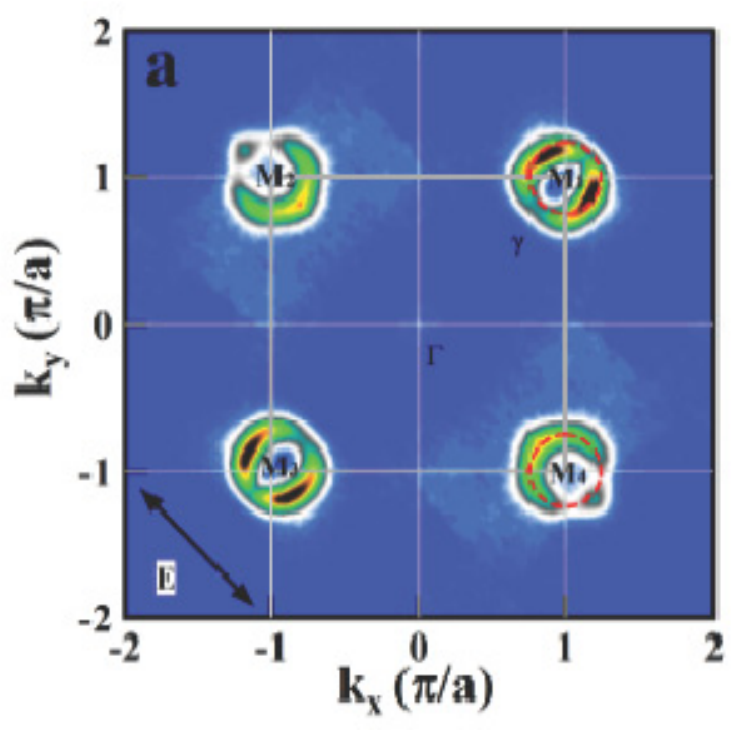}}
\caption{Experimental ARPES Fermi surface of FeSe monolayer
[\onlinecite{ARP_FS_FeSe_1,ARPES_FeSe_Nature}].}
\label{FeSe_ARPES_bands}
\end{figure}

The results of our LDA calculations [\onlinecite{Nekrasov_FeSe}] of the spectrum for the
isolated FeSe monolayer together with FeSe layer on STO substrate are shown in
Fig. \ref{fese_sto_bands}. This spectrum has the form typical for FeAs based
systems and bulk FeSe as discussed in detail above. However ARPES experiments
[\onlinecite{ARP_FS_FeSe_1, ARPES_FeSe_Nature, ARP_FS_FeSe_2}] are in striking
disagreement with these results. Actually, in FeSe monolayers on STO only
electron -- like Fermi surface sheets are observed around the $M$ -- points of the
Brillouin zone, while hole -- like sheets, centered around the $\Gamma$ -- point
(in the center of the zone), are just absent. An example of such data is shown
in Fig. \ref{FeSe_ARPES_bands} (a) [\onlinecite{ARP_FS_FeSe_1}].
Similarly to intercalated FeSe systems there is no place for ``nesting'' of
Fermi surfaces -- there are just no surfaces to ``nest''!

In order to explain this contradiction between ARPES experiments
[\onlinecite{ARP_FS_FeSe_1}] and band structure calculations reflected in the
absence of hole -- like cylinders at the $\Gamma$ -- point, one can suppose it
to be the consequence of FeSe/STO monolayer stretching due to mismatch of lattice
constants of the bulk FeSe and STO. We have studied this problem by varying the
lattice parameter $a$  and Se height $z_{Se}$ in the range $\pm 5\%$ around the
bulk FeSe parameters with the account of lattice relaxation. The conclusion was
that the changes of lattice parameters do not lead to qualitative changes of
FeSe Fermi surfaces and the hole -- cylinders in the $\Gamma$ -- point always
remain more or less intact.

However, there is another rather simple possible explanation for the absence
of hole -- like  cylinders and the observed Fermi surfaces can be obtained
assuming that the system is doped by electrons.  The Fermi level has to be moved
upwards in energy by the value of $\sim $ 0.2 - 0.25 eV,
corresponding to the doping level of 0.15 - 0.2 electron per Fe ion.

The nature of this doping, strictly speaking, is not fully understood. There
is a common belief that it is associated with the formation of oxygen vacancies
in the SrTiO$_3$ substrate (in the topmost layer of TiO$_2$),
occurring during the various technological steps used during film preparation,
such as annealing, etching, etc. It should be noted that the formation of the
electron gas at the interface with the SrTiO$_3$ is rather widely known phenomenon,
which was studied for a long time [\onlinecite{Shkl_16}]. At the same time, for
FeSe/STO system this issue was not analyzed in detail and remains unexplained
(see, however, recent Refs. [\onlinecite{Mills, Agter}].

\subsubsection{LDA+DMFT results}

\begin{figure*}[!ht]
\center{\includegraphics[width=.95\linewidth]{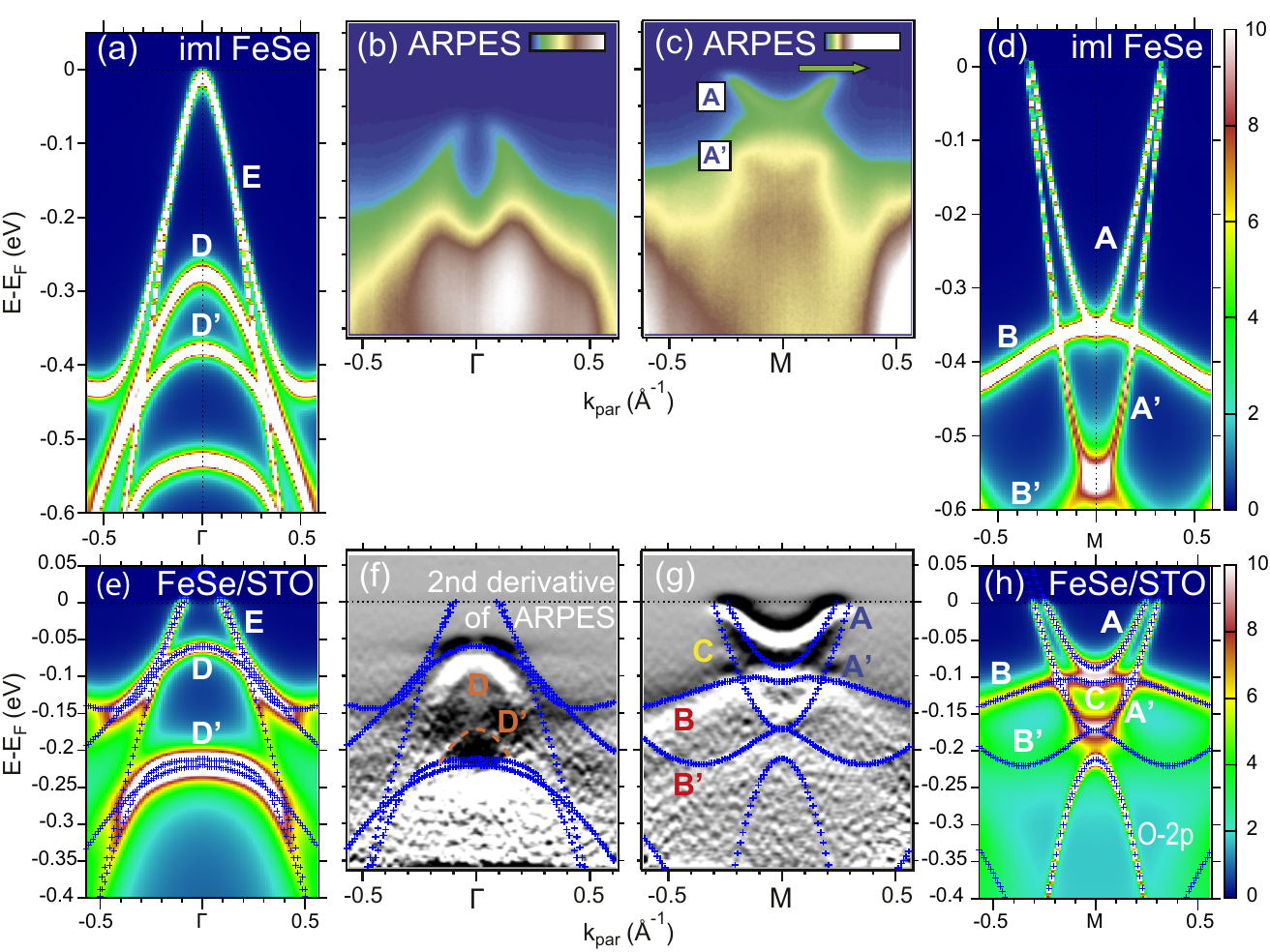}}
\caption{(a), (d) panels -- LDA+DMFT spectral function maps of isolated FeSe
monolayer [\onlinecite{FeSe_STO_2017}] and (b),(c) -- experimental ARPES data around
$\Gamma$ and M points and (f), (g) corresponding second derivatives of ARPES data
for FeSe/STO~[\onlinecite{FeSe_STO_arpes14}] with LDA+DMFT spectral function maxima shown with crosses; (e), (h) -- LDA+DMFT spectral function maps
[\onlinecite{FeSe_STO_2017}] with maxima shown with crosses for FeSe/STO.
To mark similar features of experimental and theoretical spectral function
maps $A,B,C,D,E$ letters are used (the same as in Fig.~\ref{fese_sto_bands} for
LDA bands). Fermi level is at zero energy.}
\label{fig1sto}
\end{figure*}

In Fig.~\ref{fig1sto} we compare our theoretical LDA+DMFT results
[\onlinecite{FeSe_STO_2017}], shown on panels (a,d,e,h), with experimental
ARPES data~[\onlinecite{FeSe_STO_arpes14}], shown on panels (b,c,f,g).
LDA+DMFT spectral function maps of isolated FeSe monolayer are shown in
Fig.~\ref{fig1sto}(a) and Fig.~\ref{fig1sto}(d) at $\Gamma$ and M points
respectively. For FeSe/STO LDA+DMFT calculated spectral
function maps are shown on (e), (h) panels at $\Gamma$ and M points.
The obtained LDA bandwidth $W$ of Fe-3d band in isolated FeSe monolayer
it is 5.2 eV, which is much larger than 4.3 eV obtained for FeSe/STO.
This is due to the lattice constant $a$ expanded from $a=3.765$~\AA\ to
$a=3.901$~\AA~in going from isolated FeSe monolayer to FeSe/STO (see  Fig. \ref{DOScomp}).
Thus for the same interaction strength and doping levels LDA+DMFT calculations
demonstrate substantially different band narrowing due to correlation effects.
It is a factor of 1.5 in isolated FeSe monolayer (same as bulk FeSe) and a
factor of 3 in FeSe/STO. Thus, we may conclude that FeSe/STO system is more
correlated as compared with the bulk FeSe or isolated FeSe layer with respect to $U/W$ ratio.

Most of features observed in the ARPES experiments (Fig.~~\ref{fig1sto},
panels (f),(g)) can be identified with our calculated LDA+DMFT spectral
function maps (Fig.~~\ref{fig1sto}, panels (e),(h)).
The experimental quasiparticle bands around M-point marked by $A$, $B$ and $C$
(Fig.~\ref{fig1sto}(g,h)) correspond mainly to Fe-3d$_{xz}$ and Fe-3d$_{yz}$
states, while the $A'$ and $B'$ quasiparticle bands have predominantly
Fe-3d$_{xy}$ character.

The shallow band at M-point originates from LDA Fe-3d$_{xz}$ and Fe-3d$_{yz}$
bands (see also Fig.~\ref{fese_sto_bands}) compressed by electronic correlations.
Trying to achieve the better agreement with experiments we also examined
the reasonable increase of Coulomb interaction within LDA+DMFT and the different
doping levels, but these have not produced the significant improvement of our
results. Corresponding  LDA+DMFT spectral function maps for FeSe/STO system
are presented in the Appendix with variation of only one of model parameters
$U$, $J$ or occupancy $n$ while the other two remain fixed.

The $C$ quasiparticle band near M-point appeared due to lifting of
degeneracy of Fe-3d$_{xz}$ and Fe-3d$_{yz}$ bands (which is in contrast to
isolated FeSe layer, see panel Fig.~\ref{fig1sto}(d)). The origin of this band
splitting is directly related to the $z_{Se}$ height difference below and above
Fe ions plane due to the presence of interface with SrTiO$_3$.

The appearance of $A'$ (and in some works $B'$) band in FeSe/STO is usually
attributed to forward scattering interaction with 100 meV optical
phonon of STO substrate~[\onlinecite{FeSe_STO_arpes14,Gork_1,Gork_2,Rade_1,Rade_2}].
Further in the Section \ref{sect} we will provide some estimates of such
electron-optical phonon coupling strength which in fact is obtained to be
exponentially small for the case of FeSe/STO making this scenario of the ``replica'' band formation quite questionable.
Our calculations clearly show that $A'$ band of purely
electronic nature appears almost exactly at the energies of the so called
``replica'' band with no reference to phonons. Quasiparticle masses
(as listed in Tab.~1 of the Appendix) of $A$ and $A'$ bands differ from each other
not more then by 10\%. If we concentrate our attention close to M-point the shapes
of $A$ and $A'$ bands are almost the same within the accuracy of experimental data.
Let us note here that equal shapes (or the same quasiparticle masses) of $A$ and $A'$ bands
is a keypoint of phenomenological ``replica'' band description in Refs.~[\onlinecite{FeSe_STO_arpes14,Rade_1}]
One should say here that the $B'$ band is well seen in our LDA+DMFT results (Fig.~~\ref{fig1sto}, panels (g),(h))
also without introducing of any electron-phonon coupling. 
In contrast to K$_{0.76}$Fe$_{1.72}$Se$_2$ case in FeSe/STO system
the $A'$ band is well detected in the ARPES near M-point while
near Fermi level it is strongly suppressed. This may be due to
some matrix elements effects as discussed in Refs.~[\onlinecite{NaFeSe,FeSe_arpes_dxy}] and references therein,
as well as in Refs.~[\onlinecite{DMFT2,NaFeSe}] in the context of NaFeAs
compound. Again, similar to the K$_{0.76}$Fe$_{1.72}$Se$_2$ case we propose
that $A'$ and $B'$ bands are common feature of FeSe-based materials and should be
experimentally observed irrespective of the electron-phonon scenario of the ``replica'' band.

Thus, for FeSe/STO system we observe the general agreement between
the results of LDA+DMFT calculations of Ref. [\onlinecite{FeSe_STO_2017}]
(Fig.~~\ref{fig1sto}(h))  and ARPES data~[\onlinecite{FeSe_STO_arpes14}]
(Fig.~\ref{fig1sto}(g)) on semi-quantitative level with respect to relative
positions of quasiparticle bands.  Note that the Fermi surfaces formed by the
$A$ and $A'$ bands in our LDA+DMFT calculations are nearly the same as the
Fermi surface observed at M-point by ARPES.

Actually, all quasiparticle bands in the vicinity of M-point can be well
represented as LDA bands compressed by a factor of 3 due to electronic
correlations.  This fact is clearly supported by our calculated LDA band
structure shown on Fig.~\ref{fese_sto_bands}, where different bands are marked
by letters identical to those used in Fig.~\ref{fig1sto}.

Near the M-point we also observe the O-2p$_y$ band (in the energy interval
below -0.2 eV (Fig.~\ref{fig1sto}(h)) originating from TiO$_2$ layer adjacent to
FeSe. Due to doping level used this O-2p$_y$ band goes below the Fermi level in
contrast to LDA picture shown in Fig.~\ref{fese_sto_bands} where O-2p$_y$ band
crosses the Fermi level and forms hole pocket. This observation rules out
possible nesting effects between these bands which might be expected from LDA
results~[\onlinecite{Nekrasov_FeSe}].

Now let us discuss the bands around the $\Gamma$-point, which are
shown on panels (a,b,e,f) of Fig.~\ref{fig1sto}. Here the situation is much
somehow simpler than in the case of M-point. One can see here only two bands
observed in the experiment (Fig.~\ref{fig1sto}(f)).
The $D$ quasiparticle band has predominantly Fe-3d$_{xy}$ character, while the
$D'$ quasiparticle band originates from Fe-3d$_{3z^2-r^2}$ states.
The relative locations of LDA+DMFT calculated $D$ and $D'$ bands are quite
similar to the ARPES data.

Main discrepancy of LDA+DMFT results and ARPES data here is the $E$ band shown in
Fig.~\ref{fig1sto}(e) which is not observed in the ARPES. This band corresponds
to a hybridized band of Fe-3d$_{xz}$, Fe-3d$_{yz}$ and Fe-3d$_{xy}$ states.
In principle some traces of this band can be guessed in the experimental
data of Fig.~\ref{fig1sto}(f) around -0.17 eV and near the $k$-point 0.5.
Surprisingly these are missed in the discussion of Ref.~[\onlinecite{FeSe_STO_arpes14}].
Actually, the ARPES signal from $E$ band can be weakened because of sizable
Fe-3d$_{xy}$ contribution~[\onlinecite{NaFeSe,FeSe_arpes_dxy,DMFT2,NaFeSe}] and
thus might be indistinguishable from $D$ band. Also one can imagine that for
stronger band renormalization the $E$ band becomes more flat and might merge
with $D$ band.

\begin{figure*}[!ht]
\center{\includegraphics[width=.7\linewidth]{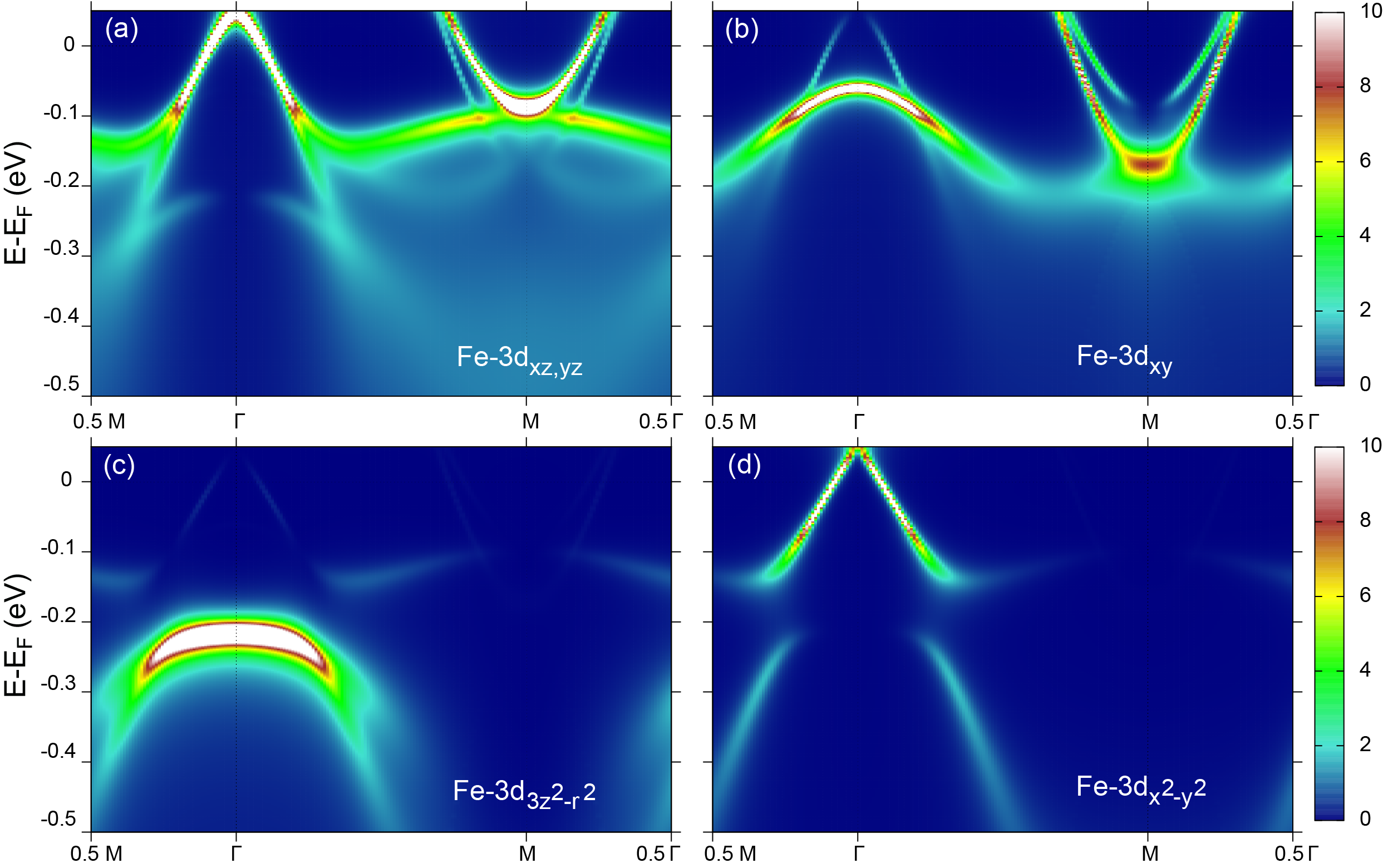}}
\center{\includegraphics[width=.7\linewidth]{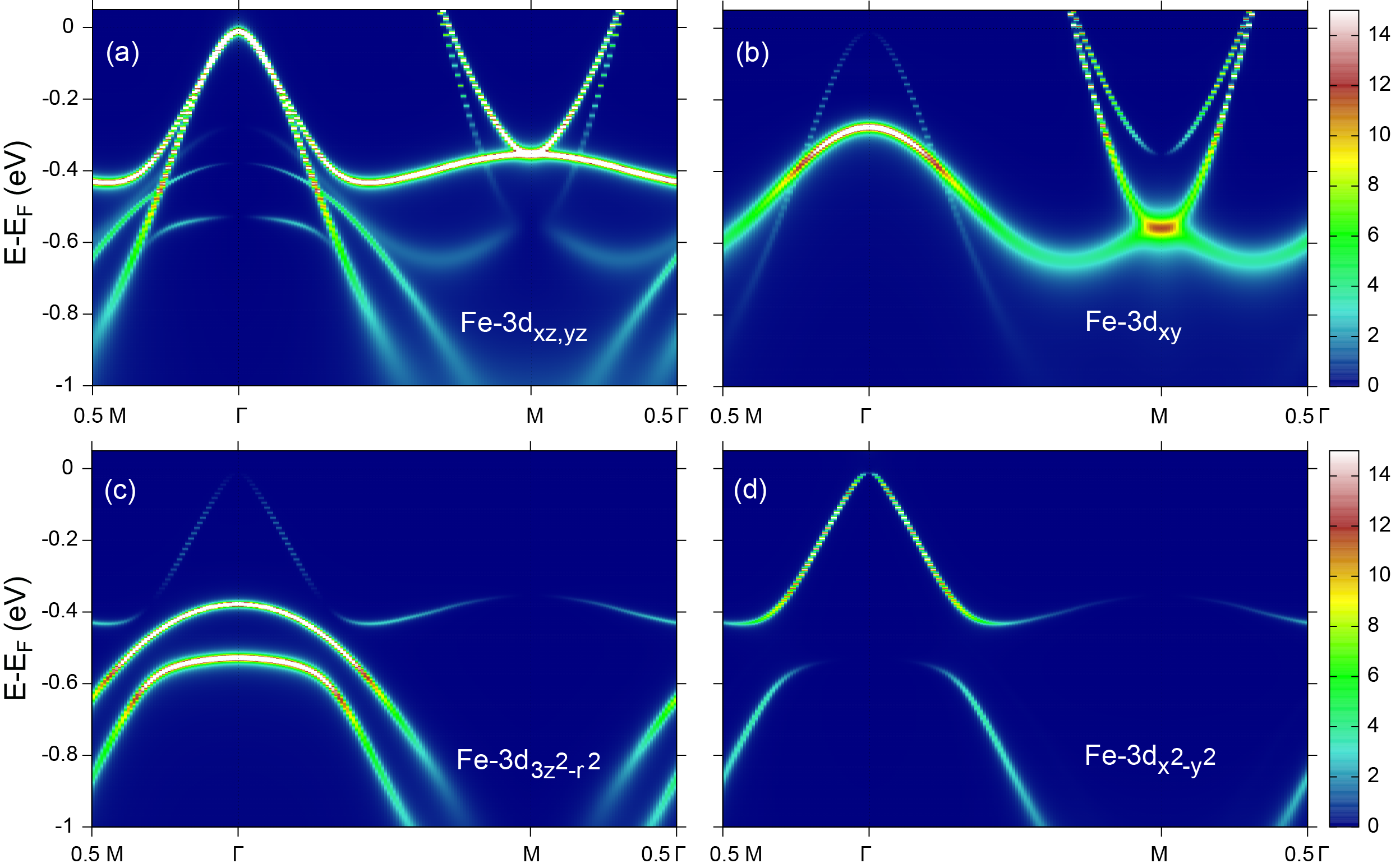}}
\caption{LDA+DMFT spectral function maps [\onlinecite{FeSe_STO_2017}]
for different Fe-3d orbitals of
FeSe monolayer on SrTiO$_3$ substrate (top) and isolated FeSe monolayer (bottom):
(a) -- Fe-3d$_{xz}$ and Fe-3d$_{yz}$, (b) -- Fe-3d$_{xy}$,
(c) -- Fe-3d$_{3z^2-r^2}$, (d) -- Fe-3d$_{x^2-y^2}$.
Fermi level is at zero energy.}
\label{or_bands_sto}
\end{figure*}

To show different Fe-3d orbitals contribution to LDA+DMFT spectral functions
of we present here the corresponding orbital resolved spectral function maps.
In Fig.~\ref{or_bands_sto} it is clearly seen that the quasiparticle bands of
isolated FeSe monolayer are well defined and have similar shape to the LDA bands
except correlation narrowing by the same constant factor for all bands.
The quasiparticle bands of FeSe/STO are more broad but still well defined.
The main contribution to spectral function near the Fermi level belongs to
Fe-3d$_{xz}$, Fe-3d$_{yz}$ and Fe-3d$_{xy}$ states both for the isolated FeSe
layer and FeSe/STO.

\section{``Replica'' band and electron -- optical phonon coupling in FeSe/STO}
\label{sect}

As we mentioned earlier, the most popular explanation of the appearance of
the ``replica'' band around the M-point in FeSe/STO is related to FeSe electrons
interaction with $\sim$100 meV optical phonons in STO. This idea was first
proposed in Ref. [\onlinecite{FeSe_STO_arpes14}], where it was experimentally observed
for the first time. In this work (see also Ref. [\onlinecite{DolgKul}]) it was also
shown that due to the peculiar nature of electron -- optical phonon interactions
at FeSe/STO interface, the appropriate coupling constant is exponentially
suppressed with transferred momentum and can be written as:
\begin{equation}
g({\bf q})=g_{0}\exp(-|{\bf q}|/q_0)
\label{g-forw}
\end{equation}
where typically $q_0\sim 0.1\frac{\pi}{a}\ll p_F$ ($a$ is the lattice constant 
and $p_F$ is the Fermi momentum), leading to the picture of nearly forward 
scattering of electrons by optical phonons.
This picture was further developed in model approach of Refs.
[\onlinecite{Rade_1,Rade_2}] where it was shown, that such coupling can also lead to
rather significant increase of the temperature of superconducting transition
$T_c$ in accordance with earlier ideas developed by Dolgov and Kuli\'c
[\onlinecite{DaDoKuO,Kulic}] (see also the review in [\onlinecite{Sad_16}]). However, the
significant effect here can be achieved only for the case of large enough
effective coupling of electrons with such forward scattering phonons.

The standard dimensionless electron -- phonon coupling constant of Eliashberg
theory for the case of optical (Einstein) phonon at FeSe/STO interface can be
written as ($N$ is the number of lattice sites) [\onlinecite{Allen_72}]:
\begin{equation}
\lambda=\frac{2}{N\Omega_0}\frac{\sum_{{\bf p},{\bf q}}|g({\bf q})|^2
\delta{(\epsilon{(\bf p)}-\mu)}\delta(\epsilon{(\bf p+q)})-\mu-\Omega_0)}
{\sum_{\bf p}\delta(\epsilon({\bf p})-\mu)},
\label{e-ph-const}
\end{equation}
where we explicitly introduced (optical) phonon frequency $\Omega_0$ in
$\delta$-function, which is usually neglected in adiabatic approximation.
In FeSe/STO system we actually have $\Omega_0>E_F$, so that it is obviously 
should be kept finite.

For simple estimates we can assume the linearized spectrum of electrons
($v_F$ is Fermi velocity):
$\xi_{p}\equiv\epsilon{(\bf p)}-\mu\approx v_F(|{\bf p}|-p_F)$
so that all calculations can be done explicitly in analytic form.
Now using (\ref{g-forw}) in (\ref{e-ph-const}) for two-dimensional case 
we can write:
\begin{widetext}
\begin{eqnarray}
\lambda=\frac{2}{\Omega_0}g_0^2\int_{-\infty}^{\infty}d\xi_{p}
\delta(\xi_{p})\int\frac{d^2q}{(2\pi)^2}\exp{\left(-\frac{2q}{q_0}\right)}
\delta(\xi_{p}-\Omega_0+v_Fq\cos\phi)=\nonumber\\
=\frac{2g_0^2}{\Omega_0}\frac{a^2}{4\pi^2}\int_{0}^{\infty}dqq
\exp\left(-\frac{2q}{q_0}\right)
\int_{0}^{2\pi}d\phi\delta(v_Fq\cos\phi-\Omega_0),
\label{lambda_int}
\end{eqnarray}
\end{widetext}
Then, after the direct calculation of all integrals, we obtain:
\begin{equation}
\lambda=\frac{g_0^2a^2}{\pi^2v_F^2}K_1\left(\frac{2\Omega_0}{v_Fq_0}\right)
\label{lambda}
\end{equation}
where $K_1(x)$ is Bessel function of imaginary argument (McDonald function).
Using the well known asymptotic behavior of $K_1(x)$ and dropping some irrelevant
constants we get:
\begin{equation}
\lambda\sim\lambda_0\frac{q_0}{4\pi p_F},
\label{lambda_0}
\end{equation}
for $\frac{\Omega_0}{v_Fq_0}\ll 1$, and
\begin{equation}
\lambda\sim\lambda_0\frac{\Omega_0}{\pi\varepsilon_F}\sqrt\frac{v_Fq_0}{\Omega_0}
\exp\left(-\frac{2\Omega_0}{v_Fq_0}\right),
\label{lambda_1}
\end{equation}
for $\frac{\Omega_0}{v_Fq_0} \gg 1$. Here we introduced the standard 
dimensionless electron -- phonon coupling constant as:
\begin{equation}
\lambda_0=\frac{2g_0^2}{\Omega_0}N(0).
\end{equation}
where $N(0)$ is the density of states at the Fermi level per one spin projection.

Now it becomes obvious that the pairing constant is exponentially suppressed
for $\frac{\Omega_0}{v_Fq_0} > 1$, which is typical for FeSe/STO interface,
where $\Omega_0>E_F\gg v_Fq_0$ [\onlinecite{Sad_16}], making the appearance of the
``replica'' band and $T_c$ enhancement due to coupling of FeSe electrons with
optical phonons of STO quite improbable. Similar conclusions were reached from
from first principles calculations of Ref. [\onlinecite{Johnson}] and the analysis of
screening of electron -- phonon interactions at FeSe/STO interface in Ref.
[\onlinecite{Mills_17}].

As we have seen above, our LDA+DMFT calculations of FeSe/STO system produced
entirely different explanation for the origin of the ``replica'' band not
related to electron -- phonon interactions.

\section{Conclusions}

Our LDA+DMFT results for FeSe monolayer materials such as K$_x$Fe$_2$Se$_2$ and
FeSe/STO provide the scenario of formation of puzzling shallow bands at the M-point
due to correlation effects on Fe-3d states only.
The detailed analysis of ARPES detected quasiparticle bands and LDA+DMFT results
shows that the closer to the Fermi level shallow band (at about 50 meV) is formed by the degenerate Fe-3d$_{xz}$ and
Fe-3d$_{yz}$ bands renormalized by correlations.
Moreover, second shallow band (at about 150 meV) can be
reasonably understood as simply correlation renormalized LDA Fe-3d$_{xy}$ band
and appears almost at the same energies as the so called ``replica'' band observed in ARPES for FeSe/STO,
usually attributed to electron interactions with optical phonons of STO. The influence of STO
substrate is reduced only to the removal of degeneracy of Fe-3d$_{xz}$ and
Fe-3d$_{yz}$ bands in the vicinity of M-point.
In the case of K$_x$Fe$_{2-y}$Se$_{2}$ most of ARPES detected bands can also be
expressed as correlation renormalized Fe-3d LDA bands.
Thus we conclude that such rather unusual band structure near Fermi level with several electron-like shallow bands is a
common feature of FeSe monolayer materials and apparently can be fully resolved in future ARPES experiments.

In principle, optical phonon mediated ``replica'' band might coincide with
demonstrated by us purely electronic shallow band if the possibility of sufficiently strong
electron-optical phonon coupling would be demonstrated. Our estimates of such coupling
strength show that it appears to be exponentially small for the FeSe/STO case.

Correlation effects alone are apparently unable to eliminate completely
the hole -- like Fermi surface at the $\Gamma$-point, which is not observed in
most ARPES experiments on FeSe/STO system.


\begin{acknowledgments}
This work was done under the State contract (FASO) No.
0389-2014-0001 and supported in part by RFBR grant No. 17-02-00015.
NSP work was also supported by the President of Russia grant for young
scientists No. Mk-5957.2016.2. The CT-QMC computations were performed at
``URAN'' supercomputer of the Institute of Mathematics and Mechanics UB RAS.

\end{acknowledgments}

\appendix*
\section{LDA+DMFT spectral function maps of FeSe/STO system for various model parameters}

In this Appendix we show the LDA+DMFT spectral function maps for FeSe/STO system
for various model parameters $U$, $J$ or occupancy $n$ while other two remains fixed.

\begin{figure*}[!ht]
\center{\includegraphics[width=.6\linewidth]{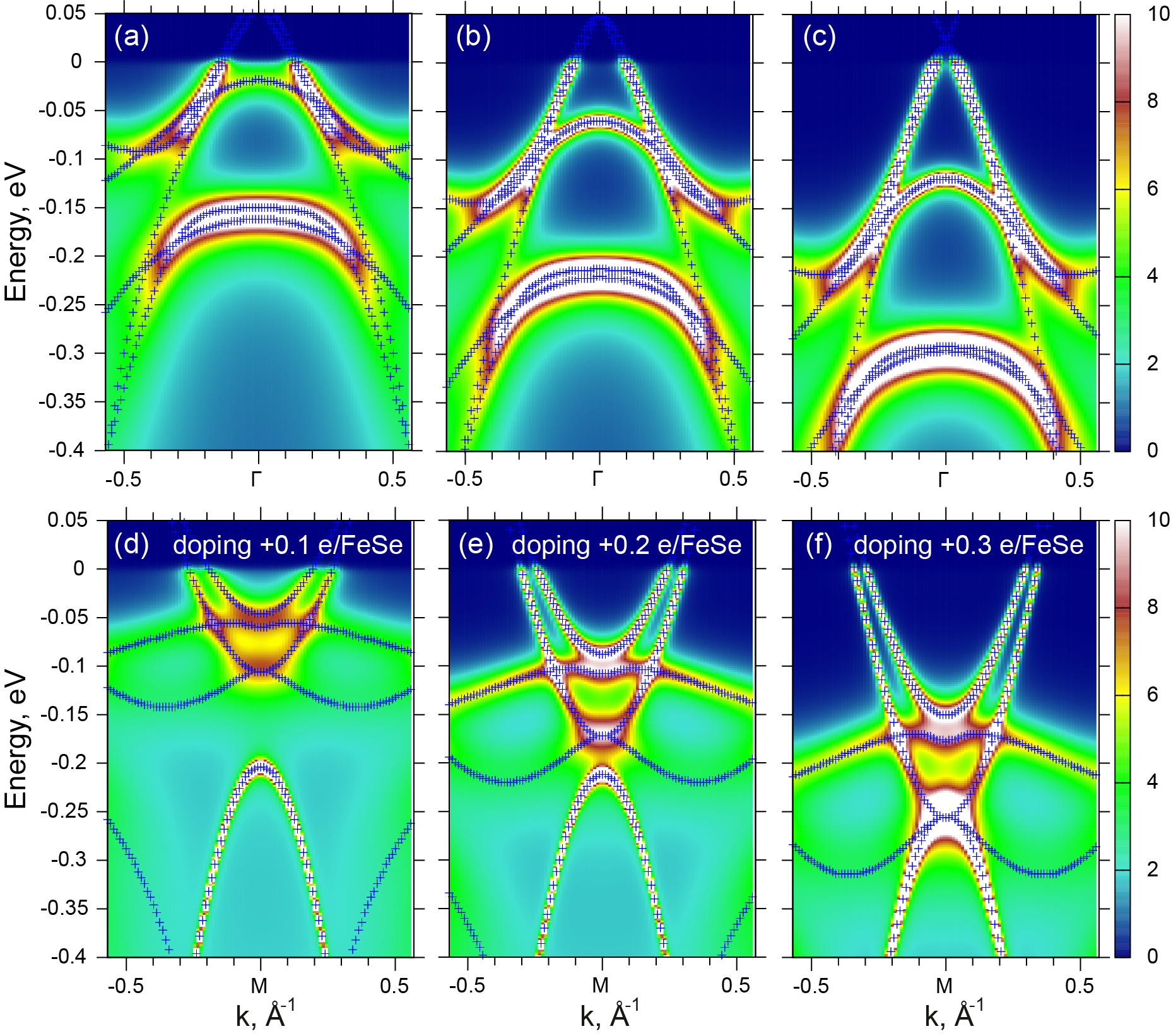}}
\caption{LDA+DMFT spectral function map of FeSe/STO for different electron doping levels (per Fe ion) for U=5.0 eV and J=0.9 eV: +0.1e, +0.2e, +0.3e (from left to right)
near $\Gamma$ (upper line) and M high symmetry points (lower line). Fermi level is at zero energy.}
\label{fese_sto_occ}
\end{figure*}

In Fig.~\ref{fese_sto_occ} we present doping dependence of the LDA+DMFT spectral function map of FeSe monolayer on SrTiO$_3$ substrate (FeSe/STO) For U=5 eV and J=0.8 eV. We assumed here three doping levels: +0.1, +0.2 (discussed in the main part of the paper) and +0.3 per Fe ion. In general such electron doping leads to a more or less rigid band shift.
However with electron doping growth the correlation strength decreases as can be seen in the upper part of the Table \ref{tab1}. Especially correlations are weakened
for t$_{2g}$ orbitals -- nearly twice weaker. It is well known behavior for iron-based
superconductors [\onlinecite{doping}]. One should note here that the doping +0.3e almost vanish Fermi surface sheets in the $\Gamma$-point (see right column of the Fig.~\ref{fese_sto_occ} on the top line) as it is observed in the ARPES (see Fig.~\ref{FSFeSeARP}. But at this doping agreement between LDA+DMFT and ARPES bands is much worth in contrast to +0.2e doping discussed above.

\begin{figure*}[!ht]
\center{\includegraphics[width=.6\linewidth]{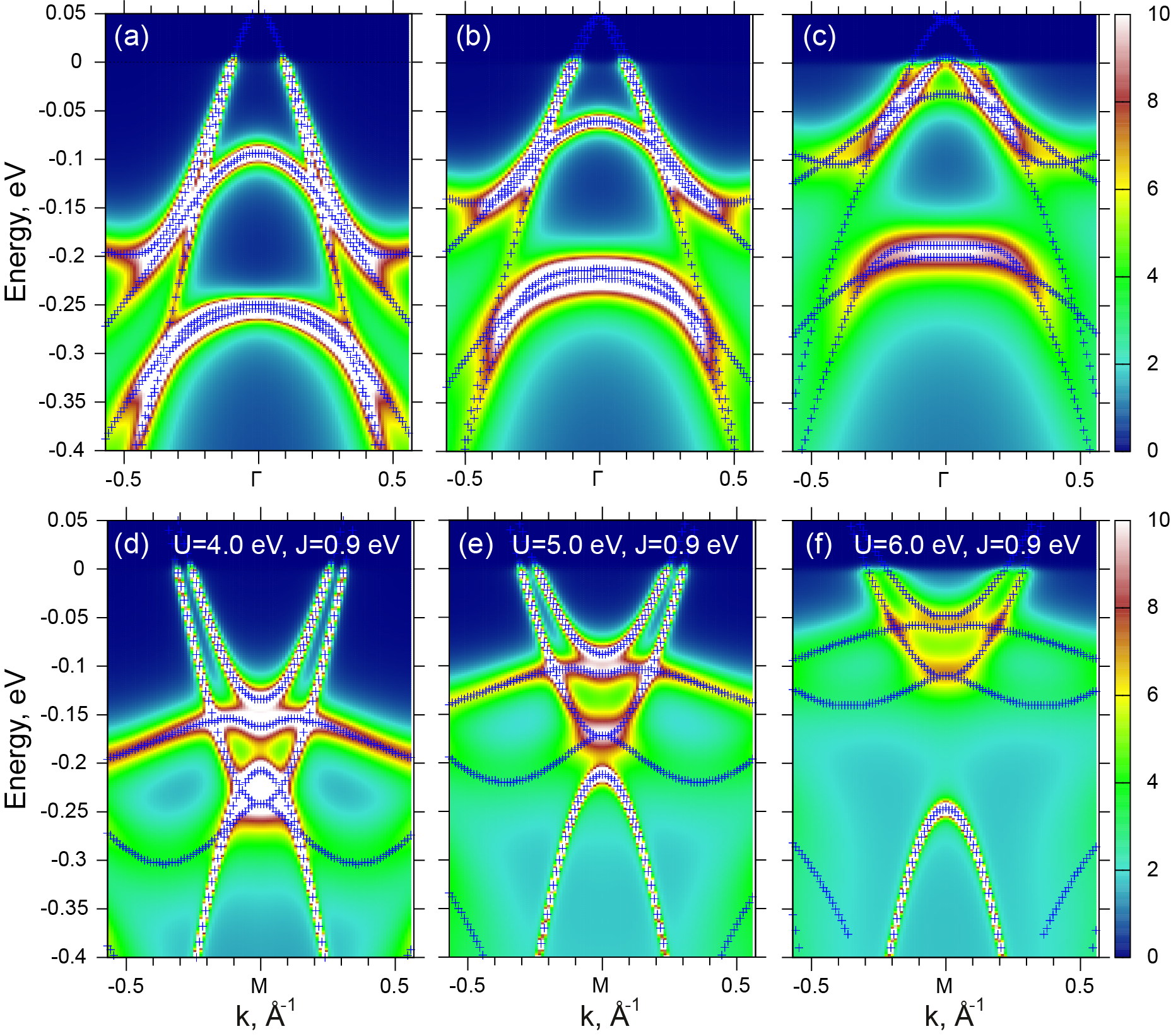}}
\caption{LDA+DMFT spectral function map of FeSe/STO for different values of U for n=+0.2e and J=0.9 eV: 4 eV, 5 eV and 6 eV (from left to right)
near $\Gamma$ (upper line) and M high symmetry points (lower line). Fermi level is at zero energy.}
\label{fese_sto_U}
\end{figure*}

The Coulomb interaction dependence of the LDA+DMFT spectral function maps of FeSe/STO is shown on Fig.~\ref{fese_sto_U}. There are three cases U=4.0 eV, U=5.0 eV and U=6.0 eV.
As it is expected increase of U gives rise to correlations (see middle part of the Table~\ref{tab1}). Such evaluation of U leads to a more less uniform bands compression. The best agreement with ARPES detected bands is found at U=5 as shown in the paper.

Perhaps the most drastic effect on LDA+DMFT results of FeSe/STO produces change of the Hund's coupling value J. In some sense it is clear since iron-based superconductors in common belief are so called ``Hund's metals'' [\onlinecite{Hundmetal}]. In Fig.~\ref{fese_sto_U} we draw Hund's coupling dependence of the LDA+DMFT spectral function map for J=0.7 eV, 0.8 eV and 0.9 eV.
In the case of J growth quasiparticle bands compression is even more evident in comparison with U evaluation (Fig.~\ref{fese_sto_U}). However mass renormalization changes approximately by a factor of 2 (see lower part of the Table \ref{tab1}) similar to those of U or n variation.

\begin{figure*}[!ht]
\center{\includegraphics[width=.6\linewidth]{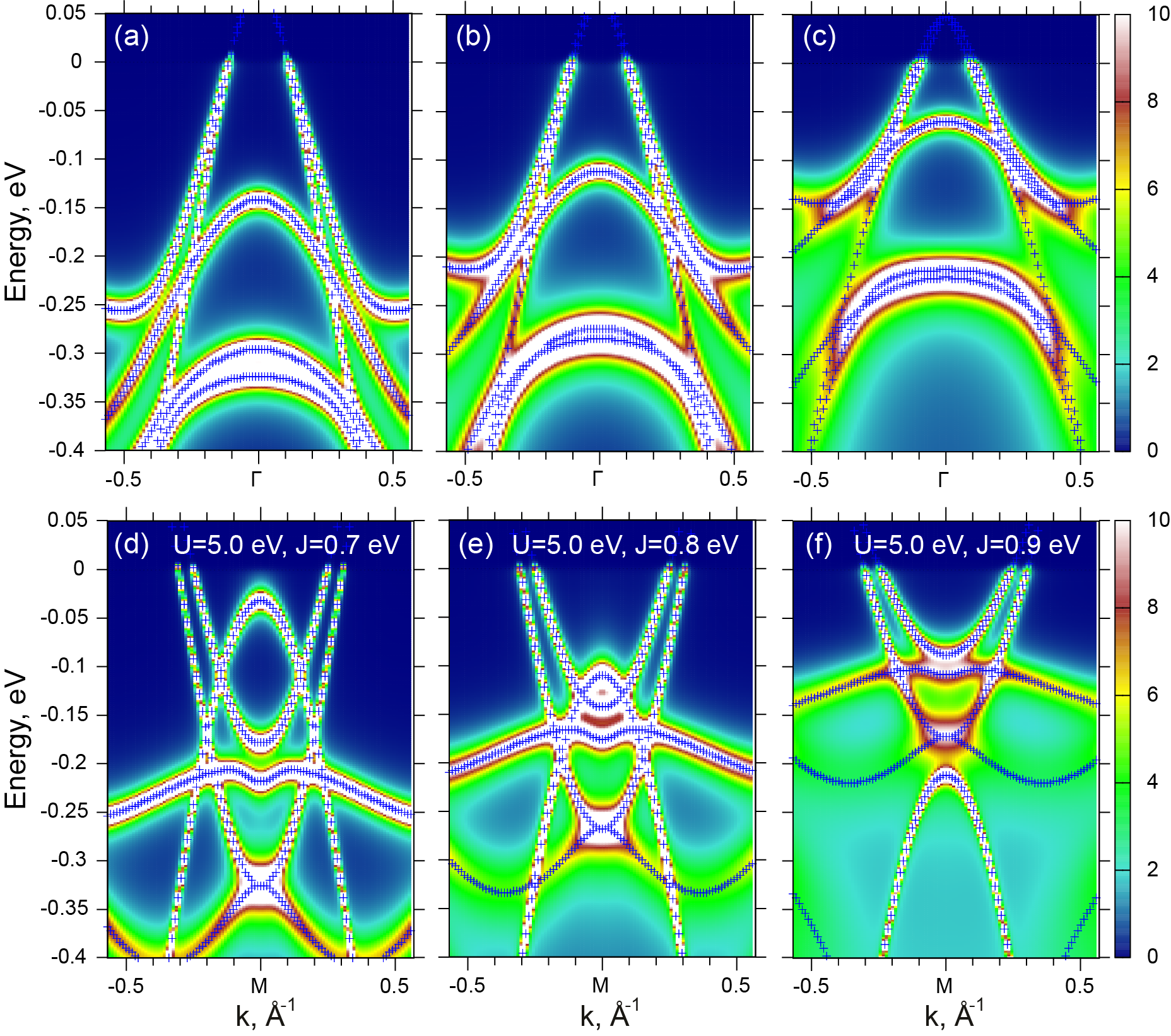}}
\caption{LDA+DMFT spectral function map of FeSe/STO for different values of J for n=+0.2e and U=5 eV: 0.7 eV, 0.8 eV and 0.9 eV (from left to right)
near $\Gamma$ (upper line) and M high symmetry points (lower line). Fermi level is at zero energy.}
\label{fese_sto_J}
\end{figure*}

\begin{table}
\caption{LDA+DMFT obtained mass renormalization values for FeSe/STO system for various model parameters
U,J and n for different Fe-3d orbitals.}
\label{tab1}
\centering
\begin{tabular}{|c|c|c|c|c|c|}
   \hline
   U=5.0 eV, J=0.9 eV(fixed) & d$_{x^2-y^2}$ & d$_{yz}$ & d$_{3z^2-r^2}$ & d$_{xz}$ & d$_{xy}$ \\
  \hline
  +0.1 e/FeSe & 2.47 & 4.25 & 2.36 & 4.04 & 4.32 \\
  +0.2 e/FeSe & 2.03 & 3.07 & 2.04 & 2.93 & 3.12 \\
  +0.3 e/FeSe & 1.84 & 2.42 & 1.83 & 2.33 & 2.47 \\
  \hline
  \hline
  +0.2 e/FeSe, J=0.9 eV (fixed) &   &   &   &   & \\
  \hline
  U=4.0 eV & 1.71 & 2.29 & 1.74 & 2.21 & 2.34 \\
  U=5.0 eV & 2.03 & 3.07 & 2.04 & 2.93 & 3.12 \\
  U=6.0 eV & 2.74 & 5.11 & 2.57 & 4.84 & 5.14 \\
  \hline
  \hline
  +0.2 e/FeSe, U=5.0 eV (fixed) &   &   &   &   & \\
  \hline
  J=0.7 eV & 1.49 & 1.73 & 1.56 & 1.69 & 1.75 \\
  J=0.8 eV & 1.63 & 2.05 & 1.70 & 1.99 & 2.08 \\
  J=0.9 eV & 2.03 & 3.07 & 2.04 & 2.93 & 3.12 \\
  \hline
\end{tabular}
\end{table}

Finally one can say that rather moderate change of model parameters for the FeSe/STO system can produce quite drastic influence on its electronic properties.

\bibliography{./CM_Review}

\end{document}